# Coherent scattering of an optically-modulated electron beam by atoms

Yuya Morimoto[1]*, Peter Hommelhoff[1] and Lars Bojer Madsen[2]†

[1] *Laser Physics, Department of Physics, Friedrich-Alexander-Universität Erlangen-Nürnberg (FAU), Staudtstraße 1, 91058, Erlangen, Germany*

[2] *Department of Physics and Astronomy, Aarhus University, 8000 Aarhus C, Denmark*

*yuya.morimoto@fau.de

†bojer@phys.au.dk

**Recent technological advances allowed the coherent optical manipulation of high-energy electron wave packets with attosecond precision. Here we theoretically investigate the collision of optically-modulated pulsed electron beams with field-free atomic targets and reveal a quantum interference associated with different momentum components of the incident broadband electron pulse, which coherently modulates both the elastic and inelastic scattering cross sections. We show that the quantum interference has a high spatial sensitivity at the level of Angstroms, offering potential applications in high-resolution ultrafast electron microscopy. Our findings are rationalized by a simple model.**

# I. INTRODUCTION

When a free electron interacts with an optical field in the presence of a third body, the electron exchanges energy and momentum with the field [1–12]. The photon-electron coupling in the vicinity of nanomaterials forms the basis of photon-induced near-field electron microscopy [7,13] and chip-scale dielectric laser accelerators [14–16]. Because electrons of different kinetic energies travel with different velocities, the temporal density of the optically-modulated beam is reshaped during the propagation in vacuum, leading to the generation of attosecond electron pulse trains [17–24].

In contrast to previously considered ultrashort electron pulses with broad Gaussian energy spectra [25–29], the energy and momentum distributions of optically-modulated electron beams consist of coherent discrete photon peaks, which might lead to unique phenomena. For example, it has been predicted that when an optically-modulated beam is employed for the excitation of a two-level system located outside the beam, the excitation probability may be enhanced when the transition energy matches an integer times the photon energy of the modulating laser beam [30–33]. Recent studies showed that this process can be described by the classical electric and magnetic fields associated with the temporal density of the modulated beam [31,32].

In this work, we investigate quantum mechanical effects in the scattering of an optically-modulated high-energy (keV to tens of keV) electron wave packet by a field-free atomic target. By using a time-dependent quantum mechanical theory [34] with projectile wave packets [35] adapted from [27,28], which is beyond the standard plane wave approximation, and taking hydrogen as an example of a target, we show that a quantum interference occurs through the coherent contributions of different momentum components of the incident beam enabled by the momentum transfer to the target. This interference modulates the elastic and inelastic cross sections by more than 30%. We also show that the strongest modulation of cross sections is induced when the target atom is located at the center of the electron beam. The sign of the modulation is reversed when the target is several Angstroms away from the beam center, which can be used for sub-nanometer imaging. We hence expect applications in attosecond imaging, atomic collisions or radiology based on the insights into the fundamental quantum mechanical interaction between the optically-shaped electrons and atoms, as laid out in this work.

The paper is organized as follows. The theoretical model is formulated in Sec. II. The results are discussed in Sec. III. Section IV draws the conclusions. The Appendix gives a detailed account of the derivations leading to the theory results of the main text.

## II. THEORETICAL MODEL

The physical system of this work is illustrated in Fig. 1(a). A non-relativistic electron wave packet propagating along the *z*-axis with group velocity $\boldsymbol{v}_e$ and central longitudinal momentum $\hbar \boldsymbol{k}_e$ is coherently accelerated or decelerated in the presence of a modulation element by an optical field of wavelength $\lambda$ and angular frequency $\omega$. As in many experiments [10,18–20,22,36], we consider a modulation field whose duration is longer than that of the electron wave packet so that the electric field amplitude ($F_0$) is constant over the entire wave packet. Examples of the modulation element include membranes [6,9,11,18–20,24,36,37], nanomaterials [3,7,10], nanofabricated dielectrics [14–16,22,23] and prisms [12,38]. Regardless of its type, the strength of the optical modulation is characterized by a dimensionless coupling parameter $g \propto eF_0 k_e/(m_e \omega^2)$, where $e$ is the unit charge and $m_e$ the electron rest mass [10,11,36,39–41]. The information on the three-body interaction between the electron, the optical field and the modulation element is captured by $g$. Larger $|g|$ yields broader energy and momentum spectra. The modulated electron wave packet collides with an atom located at $z = 0$ where the electron beam is transversally focused. The target is placed outside the laser field. We express the target position as $\boldsymbol{b} = (b_x, b_y, 0)$. We control the distance between the modulation stage and the target, $L_p = v_e t_p$, where $t_p$ is the corresponding propagation duration. Because different energy and momentum components have different phase velocities, the free-space propagation shifts the relative phase between them, which reshapes the temporal density of the electron wave packet (see below for details). The momenta of the incident and scattered electrons, $\hbar \boldsymbol{k}_i$ and $\hbar \boldsymbol{k}_f$, are described by their lengths $\hbar k_i$ and $\hbar k_f$, and polar and azimuthal angles $(\theta_i, \varphi_i)$ and $(\theta_f, \varphi_f)$. As illustrated in Fig. 1(b), quantum interference occurs when different momentum components of the incident beam, $\hbar \boldsymbol{k}_i$ and $\hbar \boldsymbol{k}_i'$, contribute to the same final momentum $\hbar \boldsymbol{k}_f$.

### A. Time-dependent perturbation theory

To consider the scattering of the broadband electron beam including its spatiotemporal structure, we adopt a time-dependent *S*-matrix formalism [27,28,34] with three-dimensional electron wave packets [35,42,43], which is beyond the standard theory using a plane wave for the asymptotic incoming state. Here we consider the scattering of an electron wave packet by an atom (A):

$$e(\boldsymbol{k}_i) + \mathrm{A}(\boldsymbol{k}_{A,i}, n) \to e(\boldsymbol{k}_f) + \mathrm{A}(\boldsymbol{k}_{A,f}, m), \tag{1}$$

where $\boldsymbol{k}_{A,i}$ and $\boldsymbol{k}_{A,f}$ are wavevectors of the target before and after the scattering, respectively. The quantum numbers $n$ and $m$ represent the initial and final electronic states of the field-free target atom. The incident electron is assumed to be non-relativistic. In order to obtain the scattering amplitude and probability, we first consider the time-dependent wavefunctions of the system before and after the scattering. The wavefunction before the scattering is given by

$$\Psi_i(\boldsymbol{x}_e, \boldsymbol{x}_A, \boldsymbol{r}, t) = \int d\boldsymbol{k}_i \int d\boldsymbol{k}_{A,i}\, a_e(\boldsymbol{k}_i) a_A(\boldsymbol{k}_{A,i})\, \chi_i(\boldsymbol{x}_e, t) \chi_{A,i}(\boldsymbol{x}_A, t) \psi_n(\boldsymbol{r}, t), \tag{2}$$

where $a_e(\boldsymbol{k}_i)$ and $a_A(\boldsymbol{k}_{A,i})$ are complex amplitudes describing the distributions over the momenta of the projectile electron and the target atom, respectively. The wavefunction $\chi_i(\boldsymbol{x}_e, t)$ is the plane wave part of the electron wavefunction,

$$\chi_i(\boldsymbol{x}_e, t) = \frac{1}{(2\pi)^{\frac{3}{2}}} \exp\left(i\boldsymbol{k}_i \cdot \boldsymbol{x}_e - \frac{iE_i t}{\hbar}\right), \tag{3}$$

with $E_i = \hbar^2 k_i^2/(2m_e)$, where $\hbar$ is the reduced Planck constant. $\boldsymbol{x}_e$ is the spatial coordinate of the incident electron beam. We note that the integral of $a_e(\boldsymbol{k}_i)\chi_i(\boldsymbol{x}_e, t)$ over $\boldsymbol{k}_i$ gives the time-dependent propagating electron wave packet in real space,

$$\psi_e(\boldsymbol{x}_e, t) = \int d\boldsymbol{k}_i\, a_e(\boldsymbol{k}_i)\chi_i(\boldsymbol{x}_e, t). \tag{4}$$

The wavefunction $\chi_{A,i}(\boldsymbol{x}_A, t)$ describes the external state of the field-free target and is given by a plane wave,

$$\chi_{A,i}(\boldsymbol{x}_A, t) = \frac{1}{(2\pi)^{\frac{3}{2}}} \exp\left(i\boldsymbol{k}_{A,i} \cdot (\boldsymbol{x}_A - \boldsymbol{b}) - \frac{iE_{A,i} t}{\hbar}\right), \tag{5}$$

with $E_{A,i} = \hbar^2 k_{A,i}^2/(2M_A)$, where $M_A$ is the mass of the target. Here $\boldsymbol{x}_A$ is the center-of-mass coordinate of the atom. The transversal component of $\boldsymbol{b}$, i.e., $b_\perp = \sqrt{b_x^2 + b_y^2}$ is the impact parameter. The wavefunction $\psi_n(\boldsymbol{r}, t)$ describes the initial electronic bound state of the target characterized by the quantum number $n$. We specify this target state as

$$\psi_n(\boldsymbol{r}, t) = \phi_n(\boldsymbol{r}) e^{-i\omega_n t}, \tag{6}$$

where $r$ is the internal spatial coordinate that denotes the set of all target electrons, $\phi_n(r)$ is the spatial part of the eigenfunction and $\hbar\omega_n$ is the eigenenergy of the state $n$. The atom is located outside of the laser field and $\psi_n(r,t)$ is the field-free eigenstate.

Second, the wavefunction of the system after the scattering is given by

$$\Psi_f(x_e, x_A, r, t) = \chi_f(x_e, t)\chi_{A,f}(x_A, t)\psi_m(r, t), \tag{7}$$

where

$$\chi_f(x_e, t) = \frac{1}{(2\pi)^{\frac{3}{2}}} \exp\left(i k_f \cdot x_e - \frac{iE_f t}{\hbar}\right), \tag{8}$$

with $E_f = \hbar^2 k_f^2/(2m_e)$ is the plane wave for the scattered electron, and

$$\chi_{A,f}(x_A, t) = \frac{1}{(2\pi)^{\frac{3}{2}}} \exp\left(i k_{A,f} \cdot (x_A - b) - \frac{iE_{A,f} t}{\hbar}\right), \tag{9}$$

describes the external state of the target with $E_{A,f} = \hbar^2 k_{A,f}^2/(2M_A)$. The internal final state of the target is expressed by

$$\psi_m(r, t) = \phi_m(r) e^{-i\omega_m t}, \tag{10}$$

where $\hbar\omega_m$ is the eigenenergy of the internal final target eigenstate $\phi_m(r)$. Within the first Born approximation and considering only the direct scattering, the transition amplitude is given by

$$T_{fi}(E_f, \hat{k}_f, k_{A,f}, b) = \iiint dx_e dx_A dr \int dt\, \Psi_f^*(x_e, x_A, r, t) V(x_e - x_A, r) \Psi_i(x_e, x_A, r, t), \tag{11}$$

where $V(x_e - x_A, r)$ is the electron-target interaction potential,

$$V(x_e - x_A, r) = -\frac{e^2 Z}{4\pi\varepsilon_0} \frac{1}{|x_e - x_A|} + \frac{e^2}{4\pi\varepsilon_0} \sum_{j=1}^{Z} \frac{1}{|x_e - x_A - r_j|}, \tag{12}$$

where $Z$ is the total number of electrons in the target, $\varepsilon_0$ is the vacuum permittivity. We here assume that the target is neutral. In Eq. (12), $r_j$ denotes the coordinates of the individual target electrons. Experimentally, the optical modulation was demonstrated for electrons of the energy of tens and hundreds of keV [3,7,9–12,14,18–23,36–38], which validates the use of the first Born approximation [44] and allows us to safely neglect the exchange scattering [45–47]. We note that the use of a perturbative approach for the description of the target-electron interaction (Born approximation) still accounts accurately for the time-dependence of the system [34], as is also familiar from theory of laser-assisted scattering [1] and strong-field ionization [48–50]. We note that a full numerical solution of the time-dependent scattering problem [25,51] might be an alternative approach but it is neither computationally attractive for the high-energy electrons propagating in three or four space-time dimensions nor necessarily given the accuracy of the Born approximation. Correction by higher-order Born terms might improve the accuracy especially for cases of lower-energy projectile electrons or heavier targets, whose discussion is beyond the scope of this work. By inserting the explicit expressions of the wavefunctions given above into Eq. (11), we obtain

$$T_{fi}(E_f, \widehat{\boldsymbol{k}}_f, \boldsymbol{k}_{A,f}, \boldsymbol{b}) = 2\pi\hbar \int d\boldsymbol{k}_i \int d\boldsymbol{k}_{A,i}\, a_e(\boldsymbol{k}_i)\, a_A(\boldsymbol{k}_{A,i})\, e^{-i\boldsymbol{k}_{A,i}\cdot\boldsymbol{b}} \delta(\boldsymbol{K}_f - \boldsymbol{K}_i)\delta(\varepsilon_f - \varepsilon_i) T_{mn}(\boldsymbol{k}_i, \boldsymbol{k}_f),$$

(13)

where $\boldsymbol{K}_f = \boldsymbol{k}_f + \boldsymbol{k}_{A,f}$, $\boldsymbol{K}_i = \boldsymbol{k}_i + \boldsymbol{k}_{A,i}$, $\varepsilon_f = E_f + E_{A,f} + \hbar\omega_m$, $\varepsilon_i = E_i + E_{A,i} + \hbar\omega_n$ and

$$T_{mn}(\boldsymbol{k}_i, \boldsymbol{k}_f) = \frac{1}{(2\pi)^3}\int d\boldsymbol{x}' \left[\int d\boldsymbol{r}\, \phi_m^*(\boldsymbol{r}) V(\boldsymbol{x}', \boldsymbol{r}) \phi_n(\boldsymbol{r})\right] \exp(i(\boldsymbol{k}_i - \boldsymbol{k}_f)\cdot \boldsymbol{x}'), \quad (14)$$

is the first Born scattering amplitude for the plane-wave incident electron, i.e., the elastic or inelastic atomic form factor for electron scattering [52]. The two delta functions in Eq. (13) represent the energy and momentum conservation of the scattering process. A detailed derivation and the explicit form of $T_{mn}$ are given in Appendix A1 and A2, respectively.

The transition amplitude in Eq. (13) is the fundamental quantity from which all observables can be constructed. If we assume that we do not resolve the final momentum of the target ($\hbar\boldsymbol{k}_{A,f}$), the scattering angle of electrons ($\widehat{\boldsymbol{k}}_f$) and the kinetic energy of the scattered electrons ($E_f$), then the total scattering probability $P_{mn}(\boldsymbol{b})$ from the target state $n$ to $m$ which is located at $\boldsymbol{b}$ is given by

$$P_{mn}(\boldsymbol{b}) = \int \frac{dP_{mn}(\boldsymbol{b})}{d\widehat{\boldsymbol{k}}_f}\, d\widehat{\boldsymbol{k}}_f, \quad (15)$$

where the differential probability $dP_{mn}(\boldsymbol{b})/d\widehat{\boldsymbol{k}}_f$ in the time-dependent $S$-matrix theory is given by [4]

$$\frac{dP_{mn}(\boldsymbol{b})}{d\widehat{\boldsymbol{k}}_f} = \int d\boldsymbol{k}_{A,f} \int dE_f \frac{(2\pi)^4 m_e^2}{\hbar^4} \left|T_{fi}(E_f, \widehat{\boldsymbol{k}}_f, \boldsymbol{k}_{A,f}, \boldsymbol{b})\right|^2. \quad (16)$$

Here we work in a regime where the kinetic energy of the electron beam is much higher than the electronic excitation energies of the target giving $|\boldsymbol{k}_f|/|\boldsymbol{k}_i| \approx 1$. The solid angle describing the propagation direction of the final wavevector of the scattered electron is given by $d\widehat{\boldsymbol{k}}_f = \sin\theta_f d\theta_f d\varphi_f$. The steps detailed in Appendix A3 lead to

$$P_{mn}(\boldsymbol{b}) = P_{mn}(\boldsymbol{b}, g) = \int \left|T_{fi}(\widehat{\boldsymbol{k}}_f, \boldsymbol{b}, g)\right|^2 d\widehat{\boldsymbol{k}}_f, \quad (17)$$

$$\left|T_{fi}(\widehat{\boldsymbol{k}}_f, \boldsymbol{b}, g)\right|^2 = P_0 \int k_i^3 dk_i \int d\widehat{\boldsymbol{k}}_i \int d\widehat{\boldsymbol{k}}_i'$$

$$\times a_e^*(k_i, \widehat{\boldsymbol{k}}_i', g) a_e(k_i, \widehat{\boldsymbol{k}}_i, g) T_{mn}^*(k_i, \widehat{\boldsymbol{k}}_i', \widehat{\boldsymbol{k}}_f) T_{mn}(k_i, \widehat{\boldsymbol{k}}_i, \widehat{\boldsymbol{k}}_f) e^{ik_i(\widehat{\boldsymbol{k}}_i - \widehat{\boldsymbol{k}}_i') \cdot \boldsymbol{b}}, \quad (18)$$

where

$$T_{mn}(k_i, \widehat{\boldsymbol{k}}_i, \widehat{\boldsymbol{k}}_f) = \frac{1}{(2\pi)^3} \int d\boldsymbol{x}' \left[ \int d\boldsymbol{r}\, \phi_m^*(\boldsymbol{r}) V(\boldsymbol{x}', \boldsymbol{r}) \phi_n(\boldsymbol{r}) \right] \exp\bigl(i(k_i \widehat{\boldsymbol{k}}_i - k_f \widehat{\boldsymbol{k}}_f) \cdot \boldsymbol{x}'\bigr), \quad (19)$$

with $k_f$ satisfiying

$$\frac{\hbar^2 k_f^2}{2m_e} = \frac{\hbar^2 k_i^2}{2m_e} - \hbar\omega_m + \hbar\omega_n. \quad (20)$$

Equation (20) reflects that a possible decrease in the final kinetic energy of the outgoing electron is accompanied by an excitation in the target atom. $P_0$ in Eq. (18) is a constant. Equation (18) contains the coherent term $a_e^*(k_i, \widehat{\boldsymbol{k}}_i', g) a_e(k_i, \widehat{\boldsymbol{k}}_i, g)$. Because of the momentum transfer to the target atom, different momentum components of the incident beam ($\hbar k_i \widehat{\boldsymbol{k}}_i'$ and $\hbar k_i \widehat{\boldsymbol{k}}_i$) can contribute to the same final momentum $\hbar \boldsymbol{k}_f$, [see Fig. 1(b)] and, accordingly, quantum interference occurs depending on the amplitudes and the relative phase of $a_e(k_i, \widehat{\boldsymbol{k}}_i', g)$ and $a_e(k_i, \widehat{\boldsymbol{k}}_i, g)$. Importantly the term $a_e^*(k_i, \widehat{\boldsymbol{k}}_i', g) a_e(k_i, \widehat{\boldsymbol{k}}_i, g)$ is absent in conventional electron scattering with plane waves but key in the case of the wave packets [35,42,43] and especially so for the optically-modulated electron beams of this study.

**B. Optically-modulated electron beam**

We consider an axially-symmetric optically-modulated electron beam whose momentum distribution is expressed by

$$a_e(k_i, \theta_i, g) = a_{e,\parallel}(k_i, \theta_i, g) a_{e,\perp}(k_i, \theta_i) e^{i\phi_{\text{prop}}(k_i, \theta_i, t_p)}. \quad (21)$$

The transversal momentum distribution $a_{e,\perp}(k_i, \theta_i)$ is given by

$$a_{e,\perp}(k_i, \theta_i) = \frac{1}{(2\pi\sigma_\perp^2)^{\frac{1}{2}}} \exp\left(-\frac{k_\perp^2}{4\sigma_\perp^2}\right), \quad (22)$$

where $k_\perp = k_i \sin\theta_i$ and $\hbar\sigma_\perp$ is the root-mean-square (rms) transversal momentum width. For simplicity, we introduce the angular width $\sigma_\theta = \sigma_\perp/k_e$. In electron microscopes, the typical value of $\sigma_\theta$ is in the range of 1-10 mrad. The rms spatial size at the focus ($z = 0$) in the transverse direction is given by $1/(2\sigma_\perp)$.

The optically-modulated longitudinal momentum distribution $a_{e,\parallel}(k_i, \theta_i, g)$ is expressed as a superposition of Gaussians of slightly different central momentum $\hbar k_e + N\hbar\delta k$ associated with the absorbed photon number $N$ [10,11,13,33,39],

$$a_{e,\parallel}(k_i, \theta_i, g) = \frac{1}{(2\pi\sigma_\parallel^2)^{\frac{1}{4}}} \sum_{N=-\infty}^{+\infty} J_N(2|g|)e^{i\phi_N} \exp\left(-\frac{(k_\parallel - k_e - N\delta k)^2}{4\sigma_\parallel^2}\right), \quad (23)$$

where $k_\parallel = k_i \cos\theta_i$, $\hbar\sigma_\parallel$ is the rms momentum width and $J_N$ is the Bessel function of the first kind. Negative $N$ corresponds to the emission of photons. The momentum shift $\hbar\delta k$ corresponds to the one-photon energy gain and is approximately given by

$$\hbar\delta k \cong \frac{m_e \omega}{k_e}. \quad (24)$$

According to classical mechanics, the maximal number of photons absorbed or emitted by the electron is $N = 2|g|$. Therefore, the maximal velocity shift is given by $\Delta v_{max} = 2|g|\hbar\delta k/m_e = 2|g|\omega/k_e$. The photon-exchange number ($N$) dependent phase $\phi_N = N \arg(-g)$ [11] is set to be zero, since this choice makes the temporal density after the optical modulation match experimental observations [18–24]. The longitudinal rms spatial width is $\sim 1/(2\sigma_\parallel)$, which corresponds to the rms temporal duration of $\sim 1/(2v_e\sigma_\parallel)$. We note that membranes [9,20,37] and nanofabricated dielectrics [22,23] are ideal modulation elements for the longitudinally directional momentum modulation. The phase $\phi_{\text{prop}}(k_i, \theta_i, t_p)$ in Eq. (21) represents the momentum-dependent phase shift due to the free-space propagation of duration of $t_p$ from the optical modulation to the target, and is given by

$$\phi_{\text{prop}}(k_i, \theta_i, t_p) = t_p\left(v_e k_\parallel - \frac{\hbar k_\parallel^2}{2m_e}\right), \quad (25)$$

see Appendix A4 for its derivation. The free-space propagation reshapes the real-space density of the electron beam. An interesting case is the density bunching into attosecond pulses occurring at [20,53]

$$t_{\text{bunch}} = \frac{v_e}{\omega \Delta v_{max}} = \frac{\hbar k_e^2}{2|g|m_e\omega^2} = \frac{m_e}{2|g|\hbar\delta k^2}. \quad (26)$$

More details of the free-space propagation and the attosecond bunching are discussed in the next section (Sec. II. C). Below, for convenience, $t_p$ is expressed in units of $t_{\text{bunch}}$.

## C. Numerical parameters and real-space density modulation

In this work, we assume a 10-keV electron beam ($v_e = 5.9 \times 10^7$ m/s, $k_e = 51$ Å$^{-1}$), with $\sigma_\parallel$ corresponding to a duration of 100 fs (full width at half maximum, FWHM) similar to an experiment [11], λ = 2 μm [22,23], and atomic hydrogen [Z = 1 in Eq. (12)] in 1s state as target, unless otherwise specified.

Before reporting results on the scattering, we briefly discuss the temporal dynamics of the optically-modulated incident electron beam. Figure 2(a) shows the real-space density of the propagating 10-keV electron wave packet, without the influence of the target atom. The five panels are the snapshots of the density $|\psi_e(\boldsymbol{x}_e, t)|^2$ given by Eqs. (4) and (21) for $t_p = t_{\text{bunch}}$, i.e., the case where the attosecond bunching occurs at the focal position of the electron beam, $\sigma_\theta = 1$ mrad and $2|g| = 5$. At $t = 0$ (left panel), the energy and momentum of the electron wave packet are modulated by an optical field. After a propagation duration of $t_p$ ($t = t_p$, middle panel), the transversal size reaches its minimum. At $t > t_p$ (two panels on the right side), the beam is diverging. Simultaneously with the transversal focusing dynamics, the longitudinal (or temporal) density modulation can also be seen. Right after the optical energy modulation ($t = 0$, left panel), the longitudinal density is still a Gaussian. However, after the free-space propagation ($t > 0$), the longitudinal density is modulated. At $t = t_p = t_{\text{bunch}}$ (middle panel), sharp peaks appear in the density.

Figure 2(b) compares the real-space temporal density of the incident electron beam $|\psi_e(\boldsymbol{x}_e, t)|^2$ of Eq. (4) at the transversal focus ($t = t_p$), i.e., at the position of the target ($z = 0$), as a function of the propagation time ($t_p$) for the cases of $2|g| = 1, 2$ and 5. In contrast to Fig. 2(a), the attosecond bunching occurs not only at the transversal focal position of the electron beam ($t_{\text{bunch}} = t_p$) but also before ($t_{\text{bunch}} < t_p$) and after ($t_{\text{bunch}} > t_p$) the focus. Following the change of the propagation time $t_p$ from the optical modulation to the target, the temporal density of the electron beam at the target changes. The sharp density peaks separated by an optical cycle (6.7 fs here) can be seen at $t_p = t_{\text{bunch}}$ in each panel. At larger $t_p$, the bunched peaks are temporally dispersed and overlap with the neighboring peaks. These overlaps induce interference among the peaks and produce the complex temporal density profile. The calculated temporal density evolutions in Fig. 2(b), especially the one in the right panel ($2|g| = 5$), are consistent with previous experimental [18–24] and theoretical [10,31,39] reports, showing the validity of the wavefunction and the propagation phase used in this study.

### III. RESULTS: MODULATION OF SCATTERING CROSS SECTIONS

**A. Cross sections at zero impact parameter**

Figure 1(c) shows the total cross sections without optical modulation $P_{mn}(\boldsymbol{b}, g = 0)$ [Eq. (17) with $g = 0$] of the elastic and inelastic [2s and 2p$_x$ final states, where $x$ axis is defined in the lab frame in Fig. 1(a)] scatterings as a function of the impact parameter $b_x$ calculated at $\sigma_\theta = 5$ mrad and $b_y = 0$. The cross sections decrease significantly with $b_x$, by a factor of $10^2$ at $b_x = 10$ nm. Therefore, we first focus on the case of $b_x = 0$, where the target is located at the center of the electron beam. Below, we discuss the modulation of the scattering probability associated with the optical shaping. For convenience, we define the amount of modulation as

$$M_{mn}(\boldsymbol{b}, g) = \frac{P_{mn}(\boldsymbol{b}, g) - P_{mn}(\boldsymbol{b}, g = 0)}{P_{mn}(\boldsymbol{b}, g = 0)}. \tag{27}$$

Hence, we compare the scattering probability of the optically-shaped electron wave packets to that without shaping. Figure 3(a) compares the modulation, $M_{mn}(\boldsymbol{b} = 0, g)$, for the three processes calculated with $2|g| = 5$ and $\sigma_\theta = 5$ mrad, as a function of the propagation time $t_p$. At zero free-space propagation ($t_p = 0$), we do not observe any modulations; see Appendix A5 for a formal proof of this property. However, we observe a clear modulation at $t_p > 0$; in particular ~20% suppression at $t_p \sim 1.7 t_{\text{bunch}}$ and ~10% enhancement at $t_p \sim 3.8 t_{\text{bunch}}$. All the three curves are qualitatively identical, suggesting that the modulation and interference appearing in Fig. 3(a) originate from the incident electron beam rather than the scattering processes. Equation (18) shows that different $\widehat{\boldsymbol{k}}_i$ and $\widehat{\boldsymbol{k}}_i'$ contribute to the same final momentum by virtue of momentum transfer to the target atom [Fig. 1(b)]. Hence quantum interference occurs when the momentum components with different incident angles $(\theta_i, \varphi_i)$ and $(\theta_i', \varphi_i')$ contribute to the cross section at the same scattering angle $(\theta_f, \varphi_f)$. A wider angular distribution, i.e., a larger momentum transfer therefore leads to stronger interference. As shown in Fig. 3(b), the dipole-allowed transition of 1s→2p$_x$ is dominated by forward scattering and shows a narrower angular distribution, which reduces the modulation contrast in Fig. 3(a) by 60%.

We now investigate the modulation dependence on the focusing angle of the electron beam ($\sigma_\theta$). Figure 3(c) compares the modulation of the total elastic cross section (1s→1s) calculated with four different angular widths, $\sigma_\theta = 1, 3, 5$ and 7 mrad. We observe less than 3% modulation at $\sigma_\theta \leq 3$ mrad, but up to 30% at 7 mrad. Figure 3(d) summarizes the results. Red circles in Fig. 3(d) show the modulation amplitudes $|M_{mn}(\boldsymbol{b} = 0, g)|$ at $t_p \sim 1.7 t_{\text{bunch}}$. At $\sigma_\theta = 0$, which corresponds to the infinitely large beam size, there is no modulation. At this limit, the transversal momentum distribution of Eq. (22) becomes $a_{e,\perp} \to \frac{1}{\sqrt{2}}\delta((k_i \sin \theta_i)^2)$. Thus only $\theta_i = \theta_i' = 0$ is allowed in Eq. (18) and accordingly the coherence and associated interference effect are lost. On the other hand, the modulation amplitudes increase exponentially up to $\sigma_\theta \approx 4$ mrad, as seen by comparison with the black dotted curve (the formula is derived below). The larger angular width allows the coupling of a wider range of $\theta_i$ and $\theta_i'$, leading to the stronger modulation.

We also investigate the dependence on the optical coupling strength $|g|$. We plot and compare $M_{mn}(\boldsymbol{b} = 0, g)$ [Eq. (27) with $\boldsymbol{b} = 0$] for the three cases of $2|g| = 1$, 2, and 5 as red circles in Figs. 4(a)-(c) at $\sigma_\theta = 5$ mrad, all for the total elastic (1s→1s) cross section. The curve of $2|g| = 1$ [Fig. 4(a)] shows a sinusoidal oscillation while the other two [Figs. 4(b),(c)] show non-sinusoidal shapes, suggesting that the modulation of $2|g| = 1$ can be described by a single sinusoidal function, however those of $2|g| = 2$ and 5 contain multiple contributions. At $2|g| = 1$ and 2, we observe only negative modulations ($M_{mn}(\boldsymbol{b} = 0, g) \leq 0$) while at $2|g| = 5$, we observe a positive modulation as well. We note that $|g|$-dependent modulations are also found in the inelastic scattering channels (not shown).

## B. Simple model

In order to understand the above results, we invoke a simple target-independent model. By considering the limit of very small $\sigma_\parallel$ ($\sigma_\parallel \ll \delta k$), by assuming $T_{mn} = 1$ (uniform scatterer) and by taking sets of $(k_i, \theta_i, \varphi_i, \theta_i', \varphi_i')$ giving dominant contributions, we obtain the following approximation to Eq. (18),

$$\left|T_{fi}^{\text{model}}(\boldsymbol{b}, g)\right|^2 = P_0^{\text{model}} \sum_{N=-\infty}^{+\infty} J_N(2|g|) \sum_{N'=-\infty}^{+\infty} J_{N'}(2|g|) \exp\left(-\frac{|N-N'|\delta k}{2k_e \sigma_\theta^2}\right)$$
$$\times e^{-i(N^2 - N'^2)\omega_{\delta k} t_p} J_0\left(\sqrt{2|N-N'|\delta k k_e} b_\perp\right), \quad (28)$$

where $P_0^{\text{model}}$ is a constant, $\omega_{\delta k} = \hbar \delta k^2/(2m_e) = 1/(4|g|t_{\text{bunch}})$, see Appendix A6 for the derivation. This model captures the essence of the coherent term $a_e^*(k_i, \hat{\boldsymbol{k}}_i', g) a_e(k_i, \hat{\boldsymbol{k}}_i, g)$ of Eq. (18) for the optically modulated electron beam. The numerical results of Eq. (28) are plotted in Figs. 3(d) and 4(a)-(c) as black curves. Even though there is no free parameter in Eq. (28), all the curves reproduce the results of the full simulations surprisingly well. Equation (28) shows that the quantum interference and the modulation of the scattering probabilities can be described by the combinations of two different photon-exchange channels with amplitude weights $J_N$ and $J_{N'}$. Since the absolute values of the Bessel functions significantly decrease for $|N|, |N'| > 2|g|$, it suffices to consider the limited range of $N$ and $N'$ (see Appendix A7 for detailed discussion). Moreover, because of the symmetry of the Bessel function, $J_{-N}(x) = (-1)^N J_N(x)$, most combinations of $J_N$ and $J_{N'}$ vanish after the sum over $N$ and $N'$ except for $|N - N'| = 0, 2, 4 \ldots$ Because the terms of $|N - N'| = 0$ are independent of $t_p$, the modulations seen in Figs. 3 and 4 are given by the terms satisfying $|N - N'| = 2, 4, \ldots$. The strength of the coupling is determined by the exponential term $\exp\left(-|N - N'|\delta k/(2k_e \sigma_\theta^2)\right)$, showing that a larger difference between $N$ and $N'$ gives a smaller contribution. The black dotted curve in Fig. 3(d) shows this exponential term with $|N - N'| = 2$ and is in good agreement with the full simulations (red circles). The deviation at large $\sigma_\theta$ is due to contributions from $|N - N'| \geq 4$.

In order to explain the observed oscillations of $M_{mn}(\boldsymbol{b} = 0, g)$ [Eq. (27) with $\boldsymbol{b} = 0$] in Fig. 4, we consider the combinations of $(N, N')$ yielding the dominant effects. At $2|g| = 1$ [Fig. 4(a)], we find them to be $(N, N') = (2,0), (-2,0), (0,2)$ and $(0, -2)$. The other combinations vanish or give negligibly small contributions, see Appendix A7 and Table 1 therein for detailed discussion. Equation (28) shows that the phase associated with the free-space propagation is proportional to $N^2 - N'^2$. The above combinations give $N^2 - N'^2 = 4$ or $-4$. Using $e^{ix} + e^{-ix} = 2\cos x$, where $x$ is a real number, $M_{mn}(\boldsymbol{b} = 0, g)$ within the model of Eq. (28) is reduced to the form of $-A + A\cos(4\omega_{\delta k} t_p)$ with $A > 0$, which explains the sinusoidal oscillation observed in Fig. 4(a) and $M_{mn}(\boldsymbol{b} = 0, g) \leq 0$ at any $t_p$. When we apply the same discussion to the cases of $2|g| = 2$ and 5 [Figs. 4(b) and (c), respectively], we find that $M_{mn}(\boldsymbol{b} = 0, g)$ is simplified to the forms of $-B_1 - B_2 + B_1\cos(4\omega_{\delta k} t_p) + B_2\cos(8\omega_{\delta k} t_p)$ and $C_1 - C_2 - C_3 - C_1\cos(8\omega_{\delta k} t_p) + C_2\cos(16\omega_{\delta k} t_p) + C_3\cos(20\omega_{\delta k} t_p)$, respectively, with real positive numbers $B_l, C_l > 0$ ($l = 1, 2, \ldots$), expressed in terms of Bessel functions, see Appendix A7 and Tables 2 and 3 therein. The largest frequencies $4\omega_{\delta k}$, $8\omega_{\delta k}$ and $20\omega_{\delta k}$ with $\omega_{\delta k} = 1/(4|g|t_{\text{bunch}})$, for $2|g| = 1$, 2 and 5, respectively, suggest the appearance of the first negative peaks at $t_p/t_{\text{bunch}} = \pi/2 = 1.6$ for all the three cases. In the case of $2|g| = 2$ [Fig. 4(b)], the signs of the two cosine functions are both positive, which give $M_{mn}(\boldsymbol{b} = 0, g) \leq 0$. On the other hand, in the case of $2|g| = 5$ [Fig. 4(c)], the term $\cos(8\omega_{\delta k} t_p)$ has a negative coefficient, which leads to $M_{mn}(\boldsymbol{b} = 0, g) > 0$. The frequency $8\omega_{\delta k}$ suggests the positive peak appearing at $t_p/t_{\text{bunch}} = 5\pi/4 = 3.9$ (see Appendix A7) which agrees with the full simulation.

## C. Cross sections at non-zero impact parameters

Next, we consider the modulation of the scattering probabilities at non-zero $b_\perp$, i.e., for a target displaced from the focus of the electron beam. According to Eq. (18), the parameter $\boldsymbol{b}$ induces a phase term of $e^{ik_i(\hat{\boldsymbol{k}}_i - \hat{\boldsymbol{k}}_i')\cdot \boldsymbol{b}}$. Its physical interpretation is illustrated in Fig. 5(a) for the case of $\boldsymbol{b} = (b_x, 0, 0)$. For an electron wave with an angle $\theta_i$, one can find the difference in the geometrical path length from the source as compared to the case of $\boldsymbol{b} = 0$ (green circle without filling), which is given by $b_x \sin\theta_i = \hat{\boldsymbol{k}}_i \cdot \boldsymbol{b}$, shown by the red arrow. This path difference corresponds to a phase shift of $\boldsymbol{k}_i \cdot \boldsymbol{b}$. We note that the same discussion is applied to electron diffraction by molecules in which case the path length difference occurs in the scattered waves. The sign and the magnitude of the coherent term $a_e^*(k_i, \hat{\boldsymbol{k}}_i', g) a_e(k_i, \hat{\boldsymbol{k}}_i, g)$ in Eq. (18) is now modified by the relative phase for $\boldsymbol{k}_i$ and $\boldsymbol{k}_i'$, that is $e^{ik_i(\hat{\boldsymbol{k}}_i - \hat{\boldsymbol{k}}_i') \cdot \boldsymbol{b}}$ in Eq. (18).

The simulation results $M_{mn}(\boldsymbol{b}, g)$ [Eq. (27)] for the elastic scattering using Eqs. (17)-(19) are shown in Fig. 5(d) with red filled circles. As in Figs. 3(a) and 4(c), we choose $2|g| = 5$, $\sigma_\theta = 5$ mrad and $t_p = 1.5 t_{\text{bunch}}$, which gives a negative modulation at $b_\perp = 0$. We observe an oscillation with $b_x$; the strongest negative peak at $b_x = 0$, the highest positive peak at around $b_x = 0.7$ nm and the second negative peak at around $b_x = 1.5$ nm. Nearly the same oscillation is observed for 1s→2s (black open squares). We therefore set $T_{mn} = 1$ (uniform scatterer) and simulate the dependence both on $b_x$ and $b_y$. The result shown in Fig. 5(b) shows a circular pattern. The vertical slice at $b_y = 0$ is shown in Fig. 5(d) as the green curve, well reproducing the results of 1s→1s (red circles) and 1s→2s (black open squares). On the other hand, at $t_p = 4 t_{\text{bunch}}$, which gives a positive modulation at $b_x = 0$ [see Fig. 4(c)], we obtain the result shown in Fig. 5(c). The radii of the circular patterns are nearly identical to those in Fig. 5(b), but the sign of the modulation is opposite. The oscillation for 1s→ 2p$_x$ [blue diamonds in Fig. 5(d)] has a longer period because the narrower angular distribution [Fig. 3(b)] gives smaller relative phase $(\hat{\boldsymbol{k}}_i - \hat{\boldsymbol{k}}_i') \cdot \boldsymbol{b}$, see discussion above.

In both the case of Figs. 5(c) and 5(d), the incoherent averaging over the target positions $b_x$ and $b_y$ gives net zero modulation. When we consider an ensemble of target atoms whose spatial distribution is given by $\rho(\boldsymbol{b})$, the total scattering probability is given by

$$P_{mn}^{\text{incoh}}(g) = \int \rho(\boldsymbol{b}) P_{mn}(\boldsymbol{b}, g) d\boldsymbol{b}. \tag{29}$$

When target atoms are uniformly distributed in the *x-y* plane, i.e, $\rho(\boldsymbol{b}) = \rho_z(b_z)$, the impact parameter dependence of Eq. (18) is expressed as

$$\iint_{-\infty}^{+\infty} e^{ik_i(\hat{\boldsymbol{k}}_i - \hat{\boldsymbol{k}}_i') \cdot \boldsymbol{b}} db_x db_y = (2\pi)^2 \delta\left(k_i(\hat{\boldsymbol{k}}_i - \hat{\boldsymbol{k}}_i')_x\right) \delta\left(k_i(\hat{\boldsymbol{k}}_i - \hat{\boldsymbol{k}}_i')_y\right), \tag{30}$$

where $(\hat{\boldsymbol{k}}_i - \hat{\boldsymbol{k}}_i')_x$ and $(\hat{\boldsymbol{k}}_i - \hat{\boldsymbol{k}}_i')_y$ are the *x* and *y* components of $\hat{\boldsymbol{k}}_i - \hat{\boldsymbol{k}}_i'$. The product of the two delta functions is equivalent to $\delta(\theta_i - \theta_i')\delta(\varphi_i - \varphi_i')$. Under this condition, the coherent term of Eq. (18) becomes $a_e^*(k_i, \theta_i', g) a_e(k_i, \theta_i, g) = |a_e(k_i, \theta_i, g)|^2$. Therefore, the scattering probability is independent on the phase of $a_e$ and no modulation occurs, i.e., $M_{mn} = 0$. The target atoms uniformly distributed along the *z*-axis, i.e., $\rho(\boldsymbol{b}) = \rho_{x,y}(b_x, b_y)$ also gives zero net modulation. We note that

$$\int_{-\infty}^{+\infty} e^{ik_i(\hat{\mathbf{k}}_i-\hat{\mathbf{k}}'_i)\cdot \mathbf{b}} db_z = 2\pi\delta\left(k_i(\hat{\mathbf{k}}_i - \hat{\mathbf{k}}'_i)_z\right), \tag{31}$$

where $(\hat{\mathbf{k}}_i - \hat{\mathbf{k}}'_i)_z$ is the z-component of $\hat{\mathbf{k}}_i - \hat{\mathbf{k}}'_i$. The delta function $\delta\left((\hat{\mathbf{k}}_i - \hat{\mathbf{k}}'_i)_z\right)$ is equivalent to $\delta(\theta_i - \theta'_i)$. Therefore, $M_{mn} = 0$ and no modulation is observed when incoherent averaging occurs. Thus, in order to observe the coherent effects shown in Fig. 5, one needs spatially fixed samples such as two-dimensional crystals or optically trapped atoms or nanoparticles.

We now investigate how the impact parameter dependence is scaled with the optical coupling strength $|g|$ and wavelength $\lambda$. To this end, we define the parameter $b_0$ which is the minimum impact parameter giving $M_{mn}(\mathbf{b}, g) = 0$, see Fig. 5(d). The simulated dependences for $|g|$ and $\lambda$ are plotted as green circles in Figs. 5(e) and (f), respectively. While there is almost no dependence of $b_0$ on $|g|$ [Fig. 5(e)], we observe a monotonic increase with $\lambda$ [Fig. 5(f)]. To understand these results, we return to the simple model and Eq (28). In Eq. (28), the $\mathbf{b}$ dependence is given solely by the term $J_0(\sqrt{2|N - N'|\delta k k_e} b_\perp)$, which is independent of $|g|$. By using $J_0(x) = 0$ at $x = 2.4$ and by recalling that the dominant contribution stems from $|N - N'| = 2$, the model predicts $b_0 = 2.4/\sqrt{4\delta k k_e} = 0.42$ nm at $\lambda = 2$ μm and for 10 keV electrons [black line in Fig. 5(e)], which agrees well with the simulation (green circles). The wavelength dependence of $\delta k = m_e \omega/(\hbar k_e) \propto 1/\lambda$ gives $b_0 \propto \sqrt{\lambda}$. In Fig. 5(f), the $\sqrt{\lambda}$ dependence (black curve) reproduces well the simulated results (green circles) at $\lambda \leq 5$ μm. The deviation between the exact results and the model at large $\lambda$, i.e., small $\delta k$, is caused by the consideration of just the dominant contributions in the simple model, see Appendix A6. The wavelength dependence can also be understood from Eq. (23) which suggests that for small wavelength, i.e., larger $\delta k$, a wider range of $\theta_i$ is required to cover different $N$ components. A larger $\theta_i$ gives a longer path length difference [Fig. 5(a)], giving stronger impact parameter dependence and smaller $b_0$. The dependence of $\delta k \propto 1/k_e$ suggests a weak dependence on the central velocity of the electron beam, see Sec. III.E below.

**D. High energy approximation**

Before concluding, we discuss two more aspects for future theoretical and experimental studies, namely, (i) an approximation which reduces computational complexity (Sec. III.D) and (ii) the generality of our findings for other electron beam parameters (Sec. III.E)

First, we introduce an approximation which speeds up the numerical evaluation of $P_{mn}(\mathbf{b}, g)$ [Eq. (17)] and $M_{mn}(\mathbf{b}, g)$ [Eq. (27)]. The results shown in Figs. 1 and 3-5 are obtained with Eqs. (17)-(19), which contain integrals over seven parameters in total. We introduce here a high-energy approximation which allows to perform the integral over $k_i$ analytically and to reduce computational time significantly. When the central energy of the electron beam is much higher than the energy bandwidth after the optical modulation, i.e., $k_e \gg \delta k$, and $k_e \gg \sigma_\parallel$, which is the case for all the reported experiments [17–24], the transversal momentum distribution $a_{e,\perp}(k_i, \theta_i)$ [Eq. (22)] and the scattering form factor $T_{mn}(k_i, \hat{\bm{k}}_i, \hat{\bm{k}}_f)$ [Eq.(19)] are nearly constant over the variation of $k_i$ within the momentum distribution of the electron beam. Therefore, they can be represented by their values at $k_i = k_e$,

$$a_{e,\perp}(k_i, \theta_i) \approx a_{e,\perp}(k_i = k_e, \theta_i), \tag{32}$$

and

$$T_{mn}(k_i, \hat{\bm{k}}_i, \hat{\bm{k}}_f) \approx T_{mn}(k_i = k_e, \hat{\bm{k}}_i, \hat{\bm{k}}_f). \tag{33}$$

Under this approximation, the differential scattering probability $|T_{fi}(\hat{\bm{k}}_f, \bm{b}, g)|^2$ in Eq. (18) can be expressed as

$$|T_{fi}(\hat{\bm{k}}_f, \bm{b}, g)|^2 = P_0 \int d\hat{\bm{k}}_i \int d\hat{\bm{k}}'_i \, a^*_{e,\perp}(k_i = k_e, \theta'_i) a_{e,\perp}(k_i = k_e, \theta_i)$$

$$\times T^*_{mn}(k_i = k_e, \hat{\bm{k}}'_i, \hat{\bm{k}}_f) T_{mn}(k_i = k_e, \hat{\bm{k}}_i, \hat{\bm{k}}_f) \sum_{N'=-\infty}^{+\infty} \sum_{N=-\infty}^{+\infty} J_{N'}(2|g|) J_N(2|g|) I_{k,N,N'}(\theta_i, \theta'_i, \bm{b}, t_p) \tag{34}$$

where

$$I_{k,N,N'}(\theta_i, \theta'_i, \bm{b}, t_p) = \int k_i^3 dk_i \exp(i k_i (\hat{\bm{k}}_i - \hat{\bm{k}}'_i) \cdot \bm{b})$$

$$\times \exp\left(i k_i v_e t_p (\cos\theta_i - \cos\theta'_i) - \frac{i\hbar k_i^2 t_p}{2 m_e}(\cos^2\theta_i - \cos^2\theta'_i)\right)$$

$$\times \exp\left(-\frac{(k_i \cos\theta_i - k_e - N\delta k)^2}{4\sigma_\parallel^2}\right) \exp\left(-\frac{(k_i \cos\theta'_i - k_e - N'\delta k)^2}{4\sigma_\parallel^2}\right). \tag{35}$$

This integral over $k_i$ can be performed analytically. Because we consider $\theta_i$ and $\theta'_i$ of the order of mrad, we can consider the Taylor expansions of $\cos\theta_i$ and $\cos\theta'_i$ and take the leading orders, $\cos\theta_i = 1 - \theta_i^2/2$ and $\cos\theta'_i = 1 - \theta_i'^2/2$. By using $k_e \gg \delta k, \sigma_\parallel$, we obtain approximately

$$I_{k,N,N'}(\theta_i, \theta'_i, \bm{b}, t_p)$$

$$\approx \frac{\sqrt{2\pi}\, \sigma_\parallel k_e^3}{\left(1 - i\sigma_\parallel^2 \frac{\hbar t_p}{2 m_e}(\theta_i^2 - \theta_i'^2)\right)^{\frac{7}{2}}}$$

$$\times \exp\left(-\frac{\{k_e(\theta_i^2 - \theta_i'^2) - 2\delta k(N' - N)\}^2}{32\sigma_\parallel^2\left(1 - i\sigma_\parallel^2 \frac{\hbar t_p}{2m_e}(\theta_i^2 - \theta_i'^2)\right)}\right) \exp(ik_e(\widehat{\mathbf{k}}_i - \widehat{\mathbf{k}}_i') \cdot \mathbf{b}). \tag{36}$$

In order to confirm the validity of this high-energy approximation, we compare in Fig. 6(a) the modulations $M_{mn}(\mathbf{b} = 0, 2|g| = 5)$ of elastic scattering calculated with Eq. (18) and Eq. (34). Because the absolute values of the Bessel functions decrease significantly for $|N|, |N'| > 2|g|$, the sums over the range of $-6 \leq N, N' \leq 6$ are sufficient to obtain results that do not change qualitatively with increasing $|N|, |N'|$, see discussion in Appendix A7 and Table 3 therein. The two curves are almost identical. In Fig. 6(b), we show the modulation at non-zero impact parameter, $M_{mn}(\mathbf{b} \neq 0, 2|g| = 5)$. The results given by the full simulation with Eq. (18) depicted by circles, squares and diamonds are well reproduced by the results with the high-energy approximation of Eq. (34) shown in lines. These results demonstrate the validity of the high-energy approximation. The high-energy approximation speeds up the computations by more than an order of magnitude compared to the evaluation of the full integrals. The good agreement facilitates its application in future works.

**E. Beam parameter dependence**

All the results so far were obtained by assuming 10-keV electrons of 100-fs duration. We here consider to which extent our findings are robust against wave packet duration and electron kinetic energy. First, we consider the dependence on the wave packet duration, i.e., the longitudinal momentum width $\hbar\sigma_\parallel$ [see Eq. (23)]. Figure 7(a) compares the modulation of elastic scattering probabilities at $\mathbf{b} = 0$ and at five different values of $\hbar\sigma_\parallel$. The corresponding FWHM initial wavepaket durations $\sqrt{2\ln 2}/(v_e\sigma_\parallel)$ are shown next to the curves. For the durations longer than 50 fs (red, black and green curves), the modulations are nearly identical. However, at 25 fs (blue curve), some deviation is observed especially at large propagation time $t_p \geq 4t_{\text{bunch}}$. This deviation can be attributed to the dispersion occurring in each $N$-photon component, which becomes stronger at shorter duration, i.e., larger $\hbar\sigma_\parallel$. At the very short duration of 10 fs (purple curve), which is comparable to the laser cycle (6.7 fs at 2 μm wavelength), significant deviation occurs already at $t_p = 2t_{\text{bunch}}$.

We then consider the dependence on the kinetic energy of the incident electron beam $\hbar^2 k_e^2/(2m_e)$. Green circles in Fig. 7(b) shows the zero-impact parameter $b_0$ (compare to Fig. 5) as a function of the electron-beam kinetic energy, calculated with $T_{mn} = 1$. At $\sigma_\theta = 5$ mrad, we observe noticeable modulations at energies only above 3 keV. Below that, the coupling of different $N$-photon components are negligibly small at the angular divergence. At the energies below 20 keV, $b_0$ is almost constant at around 0.4 nm, in good agreement with the simple model prediction (black curve). However, at energies higher than 50 keV, deviation from 0.4 nm can be seen. This is because the approximation used in the simple model Eqs. (A44)-(A45) (see Appendix A6) is not perfectly accurate at high kinetic energy, i.e., large $k_e$. The approximation is also not fully accurate for long wavelength, i.e., small $\delta k$, which can be seen in Fig. 5(f). Both cases yield $b_0$ larger than 0.4 nm but still at around 1 nm.

In short, our findings are robust over the wide range of electron-beam energy and pulse durations. Some quantitative deviations occur at high kinetic energy (>50 keV), short durations ($\leq 10$ fs) and long optical wavelength, but can be estimated accurately and simply by the approximated form of Eqs. (34) and (36).

## IV. CONCLUSION

In summary, we have investigated the scattering of an optically-modulated electron wave packet by a field-free atomic target with the time-dependent perturbative *S*-matrix approach. By virtue of the spatial focusing, the discrete longitudinal momentum components of the electron wave packet couple with each other via the scattering process and the associated coherent interference results in a modulation of the scattering probability. The sign and the amplitude of the modulations were controlled by the longitudinal and transversal momentum distributions as well as the dispersion of the projectile electron wave packets, i.e., the relative phases between different photon-exchange channels. This suggests the possibility to characterize an optically-modulated electron beam with unknown temporal and momentum structures through its scattering with a field-free target. Stronger modulations were observed in elastic scattering and dipole-forbidden inelastic scattering than in dipole-allowed inelastic scattering, which might be applied to electronic state-selective excitation. The scattering probability modulation has a strong impact parameter dependence and is hence very sensitive to the position of the target atom with respect to the focus of the electron beam. The largest enhancement (suppression) is predicted for a target at the center of the electron beam while the suppression (enhancement) occurs for a target only a few Angstroms away. Combined with the ability to control the spatial dependence of the scattering probability modulation with the optical wavelength of the electron beam modulation, the quantum interference reported here might facilitate spatially selective excitation or probe of, for example, optically trapped atoms or two-dimensional solids, or even provide an opportunity towards damage-reduced microscopy and high-resolution imaging with attosecond electron pulses.

**ACKNOWLEDGEMENTS**

This work is supported by the Gordon and Betty Moore Foundation (GBMF) through Grant No. GBMF4744 "Accelerator on a Chip International Program-ACHIP", the FAU Emerging Talents Initiative and the Danish Council for Independent Research (GrantNo.9040-00001B).

# APPENDIX

## A1. Derivation of Eq. (13)

By inserting the explicit expressions of the wavefunctions given by Eqs. (2)-(10) into Eq. (11), we obtain

$$T_{fi}(E_f, \hat{k}_f, k_{A,f}, b)$$

$$= \iiint dx_e dx_A \, dr \int dt \int dk_i \int dk_{A,i} \, a_e(k_i) a_A(k_{A,i}) \, e^{-ik_{A,i} \cdot b}$$

$$\times \chi_f^*(x_e, t)\chi_{A,f}^*(x_A, t)\psi_m^*(r, t)V(x_e - x_A, r) \, \chi_i(x_e, t)\chi_{A,i}(x_A, t)\psi_n(r, t),$$

$$= \frac{1}{(2\pi)^6} \int dk_i \int dk_{A,i} \, a_e(k_i) a_A(k_{A,i}) \, e^{-ik_{A,i} \cdot b}$$

$$\times \int dr \, \phi_m^*(r) \left( \int dx_A \int dx_e \, V(x_e - x_A, r) \exp(i(k_i - k_f) \cdot x_e) \exp(i(k_{A,i} - k_{A,f}) \cdot x_A) \right) \phi_n(r)$$

$$\times \int_{-\infty}^{+\infty} dt \, \exp\left( \frac{iE_f t}{\hbar} + \frac{iE_{A,f} t}{\hbar} - \frac{iE_i t}{\hbar} - \frac{iE_{A,i} t}{\hbar} \right) \exp\left( \frac{i\hbar\omega_m t}{\hbar} - \frac{i\hbar\omega_n t}{\hbar} \right). \quad (A1)$$

We first consider the integrals over $x_A$ and $x_e$,

$$\int dx_A \int dx_e \, V(x_e - x_A, r) \exp(i(k_i - k_f) \cdot x_e) \exp(i(k_{A,i} - k_{A,f}) \cdot x_A)$$

$$= \int dx_A \, \exp(i(k_i - k_f + k_{A,i} - k_{A,f}) \cdot x_A) \int dx' \, V(x', r) \exp(i(k_i - k_f) \cdot x'), \quad (A2)$$

where $x' = x_e - x_A$. The integral over $x_A$ gives a delta function,

$$\int dx_A \, \exp(i(k_i - k_f + k_{A,i} - k_{A,f}) \cdot x_A) = (2\pi)^3 \delta(K_f - K_i). \quad (A3)$$

This delta function represents the momentum conservation in the scattering process.

We then perform the integral over $t$ in Eq. (A1) and obtain

$$\int_{-\infty}^{+\infty} dt \, \exp\left( \frac{iE_f t}{\hbar} + \frac{iE_{A,f} t}{\hbar} - \frac{iE_i t}{\hbar} - \frac{iE_{A,i} t}{\hbar} \right) \exp\left( \frac{i\hbar\omega_m t}{\hbar} - \frac{i\hbar\omega_n t}{\hbar} \right) = 2\pi\hbar \, \delta(\varepsilon_f - \varepsilon_i). \quad (A4)$$

This delta function represents the energy conservation. By inserting Eqs. (A2)-(A3) into Eq. (A1), we obtain Eq. (13).

## A2. Atomic form factor

We here give explicit expression of the scattering form factor [$T_{mn}$, Eq. (19)] for the atomic hydrogen target. The spatial part of the eigenfunctions of the 1s, 2s, 2p$_x$ states are known analytically and given by

$$\phi_{1s}(\mathbf{r}) = \sqrt{\frac{1}{\pi}} \left(\frac{1}{a_0}\right)^{\frac{3}{2}} \exp\left(-\frac{r}{a_0}\right), \tag{A5}$$

$$\phi_{2s}(\mathbf{r}) = \sqrt{\frac{1}{\pi}} \left(\frac{1}{2a_0}\right)^{\frac{3}{2}} \left(1 - \frac{r}{2a_0}\right) \exp\left(-\frac{r}{2a_0}\right), \tag{A6}$$

$$\phi_{2p_x}(\mathbf{r}) = \sqrt{\frac{1}{\pi}} \left(\frac{1}{2a_0}\right)^{\frac{5}{2}} r \cos\theta \, \exp\left(-\frac{r}{2a_0}\right), \tag{A7}$$

respectively, where $r$ and $\theta$ are the length and the polar angle of $\mathbf{r}$ with respect to the $x$ axis in the laboratory frame [see Fig. 1(a)]. The transition amplitudes $T_{mn}$ of the elastic and inelastic scatterings are given by [52]

$$T_{1s,1s}(q) = \frac{-e^2}{8\pi^2 \varepsilon_0} \frac{a_0^2(a_0^2 q^2 + 8)}{(a_0^2 q^2 + 4)^2}, \tag{A8}$$

$$T_{1s,2s}(q) = \frac{e^2}{8\pi^2 \varepsilon_0} \frac{4\sqrt{2} a_0^2}{\left(a_0^2 q^2 + \frac{9}{4}\right)^3}, \tag{A9}$$

$$T_{1s,2p_x}(q) = \frac{e^2}{8\pi^2 \varepsilon_0} \frac{6\sqrt{2}\, i\, a_0 q_x}{q^2 \left(a_0^2 q^2 + \frac{9}{4}\right)^3}, \tag{A10}$$

where $\hbar \mathbf{q} = \hbar(\mathbf{k}_i - \mathbf{k}_f)$ is the momentum transfer, $q_x$ is the $x$-component of $\mathbf{q}$ and $a_0$ is the Bohr radius.

### A3. Derivation of Eqs. (18)

By using Eq. (13), the magnitude square of the transition amplitude $|T_{fi}(E_f, \hat{\mathbf{k}}_f, \mathbf{k}_{A,f}, \mathbf{b})|^2$ is given by

$$|T_{fi}(E_f, \hat{\mathbf{k}}_f, \mathbf{k}_{A,f}, \mathbf{b})|^2$$
$$= 4\pi^2 \hbar^2 \int d\mathbf{k}_i \int d\mathbf{k}_{A,i} \int d\mathbf{k}'_i \int d\mathbf{k}'_{A,i} \, a_e(\mathbf{k}_i)\, a_e^*(\mathbf{k}'_i)\, a_A(\mathbf{k}_{A,i})\, a_A^*(\mathbf{k}'_{A,i})\, e^{-i(\mathbf{k}_{A,i} - \mathbf{k}'_{A,i}) \cdot \mathbf{b}}$$
$$\times \delta(\mathbf{K}_f - \mathbf{K}_i)\delta(\mathbf{K}_f - \mathbf{K}'_i)\delta(\varepsilon_f - \varepsilon_i)\delta(\varepsilon_f - \varepsilon'_i) T^*_{mn}(\mathbf{k}'_i, \mathbf{k}_f) T_{mn}(\mathbf{k}_i, \mathbf{k}_f), \tag{A11}$$

where $\mathbf{k}_{A,f}$ is included in the delta functions $\delta(\mathbf{K}_f - \mathbf{K}_i)$ and $\delta(\mathbf{K}_f - \mathbf{K}'_i)$. Using the formula

$$\delta(x - y)\delta(x - z) = \delta(y - z)\delta(x - z), \tag{A12}$$

the delta functions in Eq. (A11) become

$$\delta(K_f - K_i)\delta(K_f - K'_i) = \delta(K'_i - K_i)\delta(K_f - K_i)$$
$$= \delta\big((k'_i + k'_{A,i}) - (k_i + k_{A,i})\big)\delta\big((k_f + k_{A,f}) - (k_i + k_{A,i})\big), \tag{A13}$$

and

$$\delta(\varepsilon_f - \varepsilon_i)\delta(\varepsilon_f - \varepsilon'_i) = \delta(\varepsilon'_i - \varepsilon_i)\delta(\varepsilon_f - \varepsilon_i). \tag{A14}$$

We first perform an integral over $k_{A,f}$ in Eq. (A11) with $\delta\big((k_f + k_{A,f}) - (k_i + k_{A,i})\big)$. The differential probability then becomes

$$\frac{dP_{mn}(b)}{d\widehat{k}_f} = \int dE_f \frac{(2\pi)^4 m_e^2}{\hbar^4} \big|T_{fi}(E_f, \widehat{k}_f, b)\big|^2, \tag{A15}$$

with

$$\big|T_{fi}(E_f, \widehat{k}_f, b)\big|^2$$
$$= 4\pi^2 \hbar^2 \int dk_i \int dk_{A,i} \int dk'_i \int dk'_{A,i}\, a_e(k_i)\, a_e^*(k'_i)\, a_A(k_{A,i})\, a_A^*(k'_{A,i})\, e^{-i(k_{A,i} - k'_{A,i}) \cdot b}$$
$$\times \delta\big((k'_i + k'_{A,i}) - (k_i + k_{A,i})\big)\delta(\varepsilon'_i - \varepsilon_i)\delta(\varepsilon_f - \varepsilon_i) T^*_{mn}(k'_i, k_f) T_{mn}(k_i, k_f), \tag{A16}$$

where

$$\varepsilon_f = E_f + \frac{\hbar^2 |k_i + k_{A,i} - k_f|^2}{2M_A} + \hbar\omega_m. \tag{A17}$$

Second, we perform an integral over $k'_{A,i}$ with $\delta\big((k'_i + k'_{A,i}) - (k_i + k_{A,i})\big)$ and obtain

$$\big|T_{fi}(E_f, \widehat{k}_f, b)\big|^2 = 4\pi^2 \hbar^2 \int dk_i \int dk_{A,i} \int dk'_i\, a_e(k_i)\, a_e^*(k'_i)$$
$$\times a_A(k_{A,i}) a_A^*(k_i + k_{A,i} - k'_i) e^{i(k_i - k'_i) \cdot b} \delta(\varepsilon'_i - \varepsilon_i)\delta(\varepsilon_f - \varepsilon_i) T^*_{mn}(k'_i, k_f) T_{mn}(k_i, k_f), \tag{A18}$$

where

$$\varepsilon'_i = E'_i + \frac{\hbar^2 |k_i + k_{A,i} - k'_i|^2}{2M_A} + \hbar\omega_n. \tag{A19}$$

Third, we perform an integral over $k_{A,i}$. To this end, by using $M_A \gg m_e$ and by assuming that the momentum distribution of the electron is narrow enough to satisfy $|k'_i| + |k_i| \cong 2|k_i|$, we obtain an approximated form of $\delta(\varepsilon'_i - \varepsilon_i)$ as

$$\delta(\varepsilon_i' - \varepsilon_i) = \delta\left(\frac{\hbar^2|\mathbf{k}_i'|^2}{2m_e} + \frac{\hbar^2|\mathbf{k}_i + \mathbf{k}_{A,i} - \mathbf{k}_i'|^2}{2M_A} + \hbar\omega_n - \frac{\hbar^2|\mathbf{k}_i|^2}{2m_e} - \frac{\hbar^2|\mathbf{k}_{A,i}|^2}{2M_A} - \hbar\omega_n\right)$$

$$\cong \delta\left(\frac{\hbar^2|\mathbf{k}_i'|^2}{2m_e} - \frac{\hbar^2|\mathbf{k}_i|^2}{2m_e}\right)$$

$$\cong \frac{m_e}{\hbar^2|\mathbf{k}_i|}\delta(|\mathbf{k}_i'| - |\mathbf{k}_i|), \tag{A20}$$

Similarly,

$$\delta(\varepsilon_f - \varepsilon_i) \cong \delta(E_f - E_i + \hbar\omega_m - \hbar\omega_n). \tag{A21}$$

In addition, we introduce one more approximation. Because the target atom is well localized in space, its momentum distribution $a_A$ is much wider than that of projectile electrons [27]. Then we can use the following approximation,

$$a_A^*(\mathbf{k}_i + \mathbf{k}_{A,i} - \mathbf{k}_i') \cong a_A^*(\mathbf{k}_{A,i}). \tag{A22}$$

We can now perform the integral over $\mathbf{k}_{A,i}$ in Eq. (A18),

$$|T_{fi}(E_f, \hat{\mathbf{k}}_f, \mathbf{b})|^2 = 4\pi^2\hbar^2 I_A \int d\mathbf{k}_i \int d\mathbf{k}_i' \, a_e(\mathbf{k}_i) \, a_e^*(\mathbf{k}_i') \, e^{i(\mathbf{k}_i - \mathbf{k}_i')\cdot\mathbf{b}}$$

$$\times \frac{m_e}{\hbar^2|\mathbf{k}_i|}\delta(|\mathbf{k}_i'| - |\mathbf{k}_i|)\delta(E_f - E_i + \hbar\omega_m - \hbar\omega_n)T_{mn}^*(\mathbf{k}_i', \mathbf{k}_f)T_{mn}(\mathbf{k}_i, \mathbf{k}_f), \tag{A23}$$

where we introduced the short-hand notation $I_A$ for the integral

$$I_A = \int d\mathbf{k}_{A,i}|a_A(\mathbf{k}_{A,i})|^2. \tag{A24}$$

For convenience, we use spherical coordinates in the evaluation of the integrals, that is, $\mathbf{k}_i$ is expressed by its length $k_i$ and the polar and azimuthal angles $(\theta_i, \varphi_i)$,

$$|T_{fi}(E_f, \hat{\mathbf{k}}_f, \mathbf{b})|^2$$
$$= 4\pi^2\hbar^2 I_A \int k_i^2 dk_i \int \sin\theta_i \, d\theta_i \int d\varphi_i \int k_i'^2 dk_i' \int \sin\theta_i' \, d\theta_i' \int d\varphi_i' \, a_e(k_i, \theta_i, \varphi_i) \, a_e^*(k_i', \theta_i', \varphi_i')$$

$$\times e^{i(\mathbf{k}_i - \mathbf{k}_i')\cdot\mathbf{b}}\frac{m_e}{\hbar^2 k_i}\delta(k_i' - k_i)\delta(E_f - E_i + \hbar\omega_m - \hbar\omega_n)T_{mn}^*(\mathbf{k}_i', \mathbf{k}_f)T_{mn}(\mathbf{k}_i, \mathbf{k}_f). \tag{A25}$$

We perform the integral over $k_i'$ by using the presence of $\delta(k_i' - k_i)$ and obtain

$$|T_{fi}(E_f, \hat{\mathbf{k}}_f, \mathbf{b})|^2 = 4\pi^2 m_e I_A \int k_i^3 dk_i \int \sin\theta_i \, d\theta_i \int d\varphi_i \int \sin\theta_i' \, d\theta_i' \int d\varphi_i'$$

$$\times a_e(k_i, \hat{\mathbf{k}}_i) \, a_e^*(k_i, \hat{\mathbf{k}}_i') \, e^{ik_i(\hat{\mathbf{k}}_i - \hat{\mathbf{k}}_i')\cdot\mathbf{b}} \, \delta(E_f - E_i + \hbar\omega_m - \hbar\omega_n)T_{mn}^*(k_i, \hat{\mathbf{k}}_i', \mathbf{k}_f)T_{mn}(k_i, \hat{\mathbf{k}}_i, \mathbf{k}_f), \tag{A26}$$

where $\hat{\boldsymbol{k}}_i$ and $\hat{\boldsymbol{k}}_i'$ are unit vectors along $\boldsymbol{k}_i$ and $\boldsymbol{k}_i'$, respectively. Finally, we perform the integral over $E_f$ in Eq. (A26) with $\delta(E_f - E_i + \hbar\omega_m - \hbar\omega_n)$ and obtain Eq. (18).

### A4. Propagation phase

The propagation phase $\phi_{\text{prop}}(k_i, \theta_i, t_p)$ [Eq. (25)] is obtained by the following procedure. The evolution of the real-space amplitude of the electron wave packet is given by the Fourier transform of the momentum-space amplitude [39] given by Eq. (4). For convenience we repeat it here,

$$\psi_e(\boldsymbol{x}_e, t) = \frac{1}{(2\pi\hbar)^{3/2}} \int d\boldsymbol{k}\, a_e(\boldsymbol{k}) \exp\left(i\boldsymbol{k}\cdot\boldsymbol{x}_e - \frac{iE_k t}{\hbar}\right), \tag{A27}$$

where the norm squared of $\psi_e(\boldsymbol{x}_e, t)$ gives the probability distribution of the electron in real-space, and the energy is given by $E_k = \hbar^2 k^2/(2m_e)$. We now consider the propagation of the electron wave packet over a distance of $L_p = v_e t_p = \hbar k_e t_p/m_e$, where $v_e$ is the group velocity of the electron wave packet taken to be along the $z$-axis, see Fig. 1(a). The phase term associated with the propagation can be expressed by

$$\exp\left(ik_\parallel v_e t_p - \frac{iE_k t_p}{\hbar}\right), \tag{A28}$$

where $k_\parallel = k_i \cos\theta_i$. This expression, however, needs to be modified as we will now discuss. The kinetic energy term $E_k$ contains not only the longitudinal component but also the transversal component. The transversal component changes the electron beam diameter at the position of the target depending on the value of $t_p$. In other words, depending on the value of $t_p$, the location of the transversal beam focus moves along the $z$-axis with respect to the target. In order to compensate this contribution and to obtain the beam focus at the target plane $b_z = 0$, we define the propagation phase as

$$\exp\left(i\phi_{\text{prop}}(k_i, \theta_i, t_p)\right) = \exp\left(ik_\parallel v_e t_p - \frac{iE_k t_p}{\hbar}\right)\exp\left(-\frac{iE_{k,\perp}(-t_p)}{\hbar}\right)$$
$$= \exp\left(ik_\parallel v_e t_p - \frac{iE_{k,\parallel} t_p}{\hbar}\right), \tag{A29}$$

where $E_{k,\perp} = \hbar^2 k_\perp^2/(2m_e)$, $k_\perp = k_i \sin\theta_i$ and $E_{k,\parallel} = \hbar^2 k_\parallel^2/(2m_e)$. The temporal evolution of the wave packet and the corresponding change of the temporal density calculated with $\phi_{\text{prop}}(k_i, \theta_i, t_p)$ are consistent with experiment results [18–24], see main text.

### A5. Modulation at zero propagation duration

We consider the differential scattering probability $\left|T_{fi}(\hat{\boldsymbol{k}}_f, \boldsymbol{b}, g)\right|^2$ [Eq.(18)] at $t_p = 0$. The numerical results in Figs. 3(a), (c) and 4 show that $M_{mn}(\boldsymbol{b} = 0, g) \cong 0$ [Eq.(27)] at $t_p = 0$. This suggests that even though the energy and momentum spectra of the electron beam are already broad at $t_p = 0$, some free-space propagation is required to give a non-zero modulation. With the high-energy approximation, whose accuracy is shown above, the integral $I_{k,N,N'}(\theta_i, \theta_i', \boldsymbol{b}, t_p)$ of Eq. (36) becomes

$$I_{k,N,N'}(\theta_i, \theta_i', \boldsymbol{b}, t_p = 0)$$

$$= \sqrt{2\pi}\, \sigma_\| k_e^3 \exp\left(-\frac{\{k_e(\theta_i^2 - \theta_i'^2) - 2\delta k(N' - N)\}^2}{32\sigma_\|^2}\right) \exp(ik_e(\hat{\boldsymbol{k}}_i - \hat{\boldsymbol{k}}_i') \cdot \boldsymbol{b})$$

(A30)

Equation (A30) shows that the integral $I_{k,N,N'}(\theta_i, \theta_i', \boldsymbol{b}, t_p = 0)$ depends only on the difference of $N$ and $N'$, i.e., $N' - N$, not the absolute numbers. We therefore express the above integral as $I_{k,N'-N}(\theta_i, \theta_i', \boldsymbol{b}, t_p = 0)$ and consider the sum over $N$ and $N'$ in Eq. (34),

$$\sum_{N'=-\infty}^{+\infty} \sum_{N=-\infty}^{+\infty} J_{N'}(2|g|) J_N(2|g|) I_{k,N'-N}(\theta_i, \theta_i', \boldsymbol{b}, t_p = 0)$$

$$= I_{k,0}(\theta_i, \theta_i', \boldsymbol{b}, t_p = 0) \sum_{N=-\infty}^{+\infty} J_N^2(2|g|) + \sum_{\substack{m=-\infty \\ m\neq 0}}^{+\infty} I_{k,m}(\theta_i, \theta_i', \boldsymbol{b}, t_p = 0) \sum_{N=-\infty}^{+\infty} J_{N+m}(2|g|) J_N(2|g|). \quad \text{(A31)}$$

By using $\sum_{N=-\infty}^{+\infty} J_N^2(x) = 1$ and

$$\sum_{N=-\infty}^{+\infty} J_N(x) J_{N+m}(x) = 0, \quad \text{(A32)}$$

for $m \neq 0$, Eq. (A31) becomes

$$\sum_{N'=-\infty}^{+\infty} \sum_{N=-\infty}^{+\infty} J_{N'}(2|g|) J_N(2|g|) I_{k,N'-N}(\theta_i, \theta_i', \boldsymbol{b}, t_p = 0) = I_{k,0}(\theta_i, \theta_i', \boldsymbol{b}, t_p = 0). \quad \text{(A33)}$$

We therefore obtain the approximate form of the differential scattering probably at $t_p = 0$ as

$$\left|T_{fi}(\hat{\boldsymbol{k}}_f, \boldsymbol{b}, g)\right|^2 = 4\pi^2 m_e I_A \int d\hat{\boldsymbol{k}}_i \int d\hat{\boldsymbol{k}}_i'\, a_{e,\perp}^*(k_i = k_e, \theta_i') a_{e,\perp}(k_i = k_e, \theta_i)$$

$$\times T_{mn}^*(k_i = k_e, \hat{\boldsymbol{k}}_i', \hat{\boldsymbol{k}}_f) T_{mn}(k_i = k_e, \hat{\boldsymbol{k}}_i, \hat{\boldsymbol{k}}_f) I_{k,0}(\theta_i, \theta_i', \boldsymbol{b}, t_p = 0). \quad \text{(A34)}$$

Notably, $|T_{fi}|^2$ is now independent on the optical modulation strength $|g|$. Therefore, the modulation of the scattering probability for any $\boldsymbol{b}$ is approximately zero, $M_{mn}(\boldsymbol{b}, g) \cong 0$, at $t_p = 0$.

### A6. Target independent model

Here we derive a simple model given by Eq. (28). For simplicity, we consider the case $T_{mn} = 1$. We refer to this case as the case of a uniform scatterer. Since we take $T_{mn} = 1$, scattering effects related to the detailed nature of the target are neglected. In this sense this model highlights physical effects related directly to the optically modulated electron beam. The validity of this assumption is illustrated in Fig. 5(d). We consider the integrals over $\varphi_i$ and $\varphi_i'$ in Eq. (18),

$$I_b = \int d\varphi_i \int d\varphi_i' \, e^{ik_i(\hat{k}_i - \hat{k}_i') \cdot \boldsymbol{b}}. \tag{A35}$$

For $\boldsymbol{b} = (b_x, b_y, b_z)$ and using spherical coordinates, we obtain

$$I_b = \exp(ik_i b_z (\cos\theta_i - \cos\theta_i'))$$

$$\times \iint_0^{2\pi} d\varphi_i \, d\varphi_i' \exp(-ik_i b_x (\sin\theta_i \cos\varphi_i - \sin\theta_i' \cos\varphi_i')) \exp(-ik_i b_y (\sin\theta_i \sin\varphi_i - \sin\theta_i' \sin\varphi_i')). \tag{A36}$$

By using

$$\int_0^{2\pi} e^{-iA\sin x - iB\cos x} dx = 2\pi J_0\left(\sqrt{A^2 + B^2}\right), \tag{A37}$$

we obtain

$$I_b = 4\pi^2 \exp(ik_i b_z (\cos\theta_i - \cos\theta_i')) J_0\left(k_i \sin\theta_i \sqrt{b_x^2 + b_y^2}\right) J_0\left(k_i \sin\theta_i' \sqrt{b_x^2 + b_y^2}\right). \tag{A38}$$

Equation (18) now becomes

$$|T_{fi}^{\text{model}}(\boldsymbol{b}, g)|^2 = 4\pi^2 P_0 \int k_i^3 dk_i \int \sin\theta_i \, d\theta_i \int \sin\theta_i' \, d\theta_i' \, a_e^*(k_i, \theta_i', g) a_e(k_i, \theta_i, g)$$

$$\times \exp(ik_i b_z (\cos\theta_i - \cos\theta_i')) J_0\left(k_i \sin\theta_i \sqrt{b_x^2 + b_y^2}\right) J_0\left(k_i \sin\theta_i' \sqrt{b_x^2 + b_y^2}\right). \tag{A39}$$

In order to further simplify the above equation, we focus on the sets of $(k_i, \theta_i, \theta_i')$ that maximize the values of $a_e(k_i, \theta_i, g)$ and $a_e^*(k_i, \theta_i, g)$ of Eq. (A39). Specifically, we first consider only values of $\theta_i$ and $\theta_i'$ which give zero arguments in the exponential functions in $a_{e,\|}(k_i, \theta_i, g)$ and $a_{e,\|}^*(k_i, \theta_i, g)$, that is [see Eq. (23)]

$$k_i \cos\theta_i - k_e - N\delta k = 0, \tag{A40}$$

and

$$k_i \cos\theta_i' - k_e - N'\delta k = 0. \tag{A41}$$

The coherent term in $a_e^*(k_i, \theta_i', g) a_e(k_i, \theta_i, g)$ in Eq. (A39) becomes

$a_e^*(k_i, \theta_i', g) a_e(k_i, \theta_i, g)$

$$= a_{e,\perp}^*(k_i,\theta_i')a_{e,\perp}(k_i,\theta_i)\frac{1}{(2\pi\sigma_\parallel^2)^{\frac{1}{2}}}\sum_{N'=-\infty}^{+\infty}\sum_{N=-\infty}^{+\infty}J_{N'}(2|g|)J_N(2|g|)\,e^{i\phi_{\text{prop}}(k_i,\theta_i,t_p)-i\phi_{\text{prop}}(k_i,\theta_i',t_p)}. \quad (A42)$$

By denoting the angles $\theta_i$ and $\theta_i'$ that satisfy Eqs. (A40) and (A41) $\theta_N$ and $\theta_{N'}'$, which are functions of $k_i$, the term for the propagation phase in Eq. (A42) becomes

$$e^{i\phi_{\text{prop}}(k_i,\theta_N,t_p)-i\phi_{\text{prop}}(k_i,\theta_{N'}',t_p)}$$

$$=\exp\left(ik_i\cos\theta_N\,v_e\,t_p - \frac{i\hbar(k_i\cos\theta_N)^2}{2m_e}t_p\right)\exp\left(-ik_i\cos\theta_{N'}'\,v_e\,t_p + \frac{i\hbar(k_i\cos\theta_{N'}')^2}{2m_e}t_p\right)$$

$$=\exp(-i(N^2-N'^2)\omega_{\delta k}\,t_p), \quad (A43)$$

As a second approximation, we only consider values of $k_i$ which give zero arguments in the exponential function in $a_{e,\perp}$ or $a_{e,\perp}^*$, that is, $\sin\theta_N = 0$ ($N' \le N$) or $\sin\theta_{N'}' = 0$ ($N' \ge N$). In the case of $\sin\theta_N = 0$, we obtain from Eqs. (A40) and (A41),

$$k_i = k_e + N\delta k, \quad (A44)$$

$$k_i\cos\theta_{N'}' = k_e + N'\delta k, \quad (A45)$$

and $a_{e,\perp}^*(k_i,\theta_i')a_{e,\perp}(k_i,\theta_i)$ in Eq. (A42) becomes

$$a_{e,\perp}^*(k_i)a_{e,\perp}(k_i) = \exp\left(-\frac{(k_i\sin\theta_{N'}'(k_i))^2}{4\sigma_\perp^2}\right) \approx \exp\left(-\frac{(N-N')\delta k}{2k_e\sigma_\theta^2}\right), \quad (A46)$$

where $\delta k^2$ is neglected. The term for the impact parameter dependence $I_b$ of Eq.(A38) becomes

$$I_b = 4\pi^2\exp(ik_ib_z(\cos\theta_N-\cos\theta_{N'}'))J_0(0)\,J_0\left(k_i\sin\theta_{N'}'\sqrt{b_x^2+b_y^2}\right)$$

$$\approx 4\pi^2\exp(ib_z(N-N')\delta k)J_0\left(\sqrt{2(N-N')\delta k\,k_e}\sqrt{b_x^2+b_y^2}\right). \quad (A47)$$

On the other hand, in the case of $\sin\theta_{N'}' = 0$ ($N' \ge N$), we obtain from Eqs. (A40) and (A41) and following the procedure in Eqs. (A44)-(A47),

$$a_{e,\perp}^*(k_i)a_{e,\perp}(k_i) \approx \exp\left(-\frac{(N'-N)\delta k}{2k_e\sigma_\theta^2}\right), \quad (A48)$$

and

$$I_b = 4\pi^2\exp(ik_ib_z(\cos\theta_i-\cos\theta_i'))J_0\left(k_i\sin\theta_i\sqrt{b_x^2+b_y^2}\right)J_0(0)$$

$$\approx 4\pi^2 \exp(ib_z(N-N')\delta k) J_0\left(\sqrt{2(N'-N)\delta k k_e}\sqrt{b_x^2+b_y^2}\right). \tag{A49}$$

In total, we obtain Eq. (28). For convenience, we repeat it here,

$$\left|T_{fi}^{\text{model}}(\boldsymbol{b},g)\right|^2 = P_0^{\text{model}} \times \sum_{N=-\infty}^{+\infty} J_N(2|g|) \sum_{N'=-\infty}^{+\infty} J_{N'}(2|g|) \exp\left(-i(N^2-N'^2)\omega_{\delta k}\, t_p\right)$$
$$\times \exp\left(-\frac{|N-N'|\delta k}{2k_e\sigma_\theta^2}\right) \exp(ib_z(N-N')\delta k) J_0\left(\sqrt{2|N-N'|\delta k k_e}\sqrt{b_x^2+b_y^2}\right). \tag{A50}$$

where $P_0^{\text{model}}$ is a constant. This equation shows that the scattering probability can be estimated by considering sums of products of $J_N$ and $J_{N'}$, reflecting different photon-exchange channels. The strength of the coupling between different channels is given by the term of $\exp\left(-\frac{|N-N'|\delta k}{2k_e\sigma_\theta^2}\right)$. The propagation effect, i.e., the dependence on $t_p$, is given by $\exp\left(-i(N^2-N'^2)\omega_{\delta k}\, t_p\right)$. The impact parameter dependence is given by $I_b = 4\pi^2 \exp(ib_z(N-N')\delta k) J_0\left(\sqrt{2|N-N'|\delta k k_e}\sqrt{b_x^2+b_y^2}\right)$.

In the case of no optical modulation, the laser-electron coupling vanishes, $2|g| = 0$, and

$$\left|T_{fi}^{\text{model}}(\boldsymbol{b},g=0)\right|^2 = P_0^{\text{model}} \times \sum_{N=-\infty}^{+\infty} |J_N(0)|^2 = P_0^{\text{model}}. \tag{A51}$$

The modulation of the scattering probability is therefore given by

$$M_{mn}^{\text{model}}(\boldsymbol{b},g) = \frac{\left|T_{fi}^{\text{model}}(\boldsymbol{b},g)\right|^2 - \left|T_{fi}^{\text{model}}(\boldsymbol{b},g=0)\right|^2}{\left|T_{fi}^{\text{model}}(\boldsymbol{b},g=0)\right|^2}$$

$$= -1 + \sum_{N=-\infty}^{+\infty} J_N(2|g|) \sum_{N'=-\infty}^{+\infty} J_{N'}(2|g|) \exp\left(-i(N^2-N'^2)\omega_{\delta k}\, t_p\right)$$
$$\times \exp\left(-\frac{|N-N'|\delta k}{2k_e\sigma_\theta^2}\right) \exp(ib_z(N-N')\delta k) J_0\left(\sqrt{2|N-N'|\delta k k_e}\sqrt{b_x^2+b_y^2}\right). \tag{A52}$$

We next consider $\left|T_{fi}^{\text{model}}(\boldsymbol{b},g)\right|^2$ and $M_{mn}^{\text{model}}(\boldsymbol{b},g)$ at zero propagation duration $t_p = 0$. We can rewrite Eq. (A50) as

$$\left|T_{fi}^{\text{model}}(\boldsymbol{b},g)\right|^2/P_0^{\text{model}} = \sum_{N=-\infty}^{+\infty} J_N(2|g|) \sum_{N'=-\infty}^{+\infty} J_{N'}(2|g|) f(N-N',\boldsymbol{b})$$

$$= \sum_{N=-\infty}^{+\infty} J_N^2(2|g|) + \sum_{N'\neq N}\sum_{N=-\infty}^{+\infty} J_N(2|g|) J_{N'}(2|g|) f(N-N',\boldsymbol{b})$$

$$= \sum_{N=-\infty}^{+\infty} J_N^2(2|g|) + \sum_{\substack{m=-\infty \\ m \neq 0}}^{+\infty} f(m, \mathbf{b}) \sum_{N=-\infty}^{+\infty} J_N(2|g|) J_{N-m}(2|g|) \tag{A53}$$

where $f(N - N', \mathbf{b}) = \exp\left(-\frac{|N-N'|\delta k}{2k_e \sigma_\theta^2}\right) \exp(ib_z(N - N')\delta k) \, J_0\left(\sqrt{2|N - N'|\delta k k_e}\sqrt{b_x^2 + b_y^2}\right)$ and $f(0, \mathbf{b}) = 1$. By using $\sum_{N=-\infty}^{+\infty} J_N^2(x) = 1$ and Eq. (A32) for $m \neq 0$, we obtain

$$\left|T_{fi}^{\text{model}}(\mathbf{b}, g)\right|^2 = P_0^{\text{model}}, \tag{A54}$$

and

$$M_{mn}^{\text{model}}(\mathbf{b}, g) = 0, \tag{A55}$$

at $t_p = 0$. Therefore, the simple target independent model yields zero modulation at $t_p = 0$, which is consistent with the scattering theory, see Section A5.

**A7. Explicit examples of the simple model**

Here we consider a few examples using the simple model above. For simplicity, we limit our considerations to the cases of $\mathbf{b} = 0$, where the $\mathbf{b}$-dependent scattering probabilities attain their maxima [see main text and Fig. 1(c)]. Equation (A50) is simplified to

$$\left|T_{fi}^{\text{model}}(\mathbf{b} = 0, g)\right|^2$$

$$= \sum_{N=-\infty}^{+\infty} J_N(2|g|) \sum_{N'=-\infty}^{+\infty} J_{N'}(2|g|) \, \exp\left(-i(N^2 - N'^2)\omega_{\delta k} \, t_p\right) \exp\left(-\frac{|N - N'|\delta k}{2k_e \sigma_\theta^2}\right), \tag{A56}$$

where the constant $P_0^{\text{model}}$ in Eq. (A50) is set to one, because we are interested only in the relative quantities and the modulation in Eq. (A52) is given by

$$M_{mn}^{\text{model}}(\mathbf{b} = 0, g) = \left|T_{fi}^{\text{model}}(\mathbf{b} = 0, g)\right|^2 - 1. \tag{A57}$$

The black curves in Fig. 3(d) and 4 are calculated using Eq. (A57).

First, at $2|g| = 1$ [Fig. 4(a)], because of the symmetry of the Bessel function $J_{-n}(x) = (-1)^n J_n(x)$, many of the combinations of $N$ and $N'$ cancel with each other, see Table 1. Since the absolute values of the Bessel functions significantly decrease for $|N| > 2|g|$, it suffices to consider the range of $-2 \leq N, N' \leq 2$. In this case we obtain

$$\left|T_{fi}^{\text{model}}(\mathbf{b} = 0, 2|g| = 1)\right|^2$$

$$\approx \sum_{N=-2}^{+2} |J_N(1)|^2 + 4J_0(1)J_2(1) \exp\left(-\frac{\delta k}{k_e \sigma_\theta^2}\right)(\exp(-i4\omega_{\delta k} t_p) + \exp(i4\omega_{\delta k} t_p))$$

$$= \sum_{N=-2}^{+2} |J_N(1)|^2 + 8 J_0(1)J_2(1) \exp\left(-\frac{\delta k}{k_e \sigma_\theta^2}\right)\cos(4\omega_{\delta k} t_p), \tag{A58}$$

Because we here consider the range of $-2 \leq N, N' \leq 2$, Eq. (A58) does not satisfy Eqs. (A54) and (A55). In order to be consistent with these equations, we take only the modulation term, i.e., the second term of Eq. (A58) and assume $M_{mn}^{\text{model}}(\boldsymbol{b} = 0, 2|g| = 1) = 0$ at $t_p = 0$. We then obtain

$$M_{mn}^{\text{model}}(\boldsymbol{b} = 0, 2|g| = 1) = -A + A \cos(4\omega_{\delta k} t_p), \tag{A59}$$

where $A = 8 J_0(1)J_2(1) \exp\left(-\frac{\delta k}{k_e \sigma_\theta^2}\right) > 0$. This equation suggests that the modulation $M_{mn}(\boldsymbol{b} = 0, 2|g| = 1)$ oscillates sinusoidally with the propagation time $t_p$ and with a period given by $2\pi/(4\omega_{\delta k}) = \pi\, t_{\text{bunch}}$. Therefore, the first negative peak in the modulation $M_{mn}^{\text{model}}(\boldsymbol{b} = 0, 2|g| = 1)$ is expected to appear at $t_p = \pi\, t_{\text{bunch}}/2 = 1.6 t_{\text{bunch}}$, which is consistent with the result of the full quantum simulation in Fig. 4(a).

Next, at $2|g| = 2$ [Fig. 4(b)], when we consider the combinations of $(N, N')$ in the range of $-3 \leq N, N' \leq 3$, there are three combinations giving non-zero contributions (see Table 2),

$$\left|T_{fi}^{\text{model}}(\boldsymbol{b} = 0, 2|g| = 2)\right|^2 \approx \sum_{N=-3}^{+3} |J_N(2)|^2 + 8 J_0(2)J_2(2) \exp\left(-\frac{\delta k}{k_e \sigma_\theta^2}\right)\cos(4\omega_{\delta k} t_p)$$

$$+ 8 J_1(2)J_3(2) \exp\left(-\frac{\delta k}{k_e \sigma_\theta^2}\right)\cos(8\omega_{\delta k} t_p) + 8 J_{-1}(2)J_3(2) \exp\left(-\frac{2\delta k}{k_e \sigma_\theta^2}\right)\cos(8\omega_{\delta k} t_p). \tag{A60}$$

However, when $\sigma_\theta$ is not very large and $\exp\left(-\frac{\delta k}{k_e \sigma_\theta^2}\right) \gg \exp\left(-\frac{2\delta k}{k_e \sigma_\theta^2}\right)$ is satisfied, which is the case at $\sigma_\theta = 5$ mrad used in this work, we can neglect the last term and obtain

$$\left|T_{fi}^{\text{model}}(\boldsymbol{b} = 0, 2|g| = 2)\right|^2$$

$$\approx \sum_{N=-3}^{+3} |J_N(2)|^2 + 8 \exp\left(-\frac{\delta k}{k_e \sigma_\theta^2}\right)\{J_0(2)J_2(2)\cos(4\omega_{\delta k} t_p) + J_1(2)J_3(2) \cos(8\omega_{\delta k} t_p)\}, \tag{A61}$$

and

$$M_{mn}^{\text{model}}(\boldsymbol{b} = 0, 2|g| = 2) = -B_1 - B_2 + B_1 \cos(4\omega_{\delta k} t_p) + B_2 \cos(8\omega_{\delta k} t_p), \tag{A62}$$

with $B_1 = 8 J_0(2)J_2(2) \exp\left(-\frac{\delta k}{k_e \sigma_\theta^2}\right) > 0$ and $B_2 = 8 J_1(2)J_3(2)\exp\left(-\frac{\delta k}{k_e \sigma_\theta^2}\right) > 0$. The faster oscillation period is $2\pi/(8\omega_{\delta k}) = \pi\, t_{\text{bunch}}$. Because the signs of the two cosine functions $B_1$ and $B_2$ are both positive, the modulation $M_{mn}^{\text{model}}(\boldsymbol{b} = 0, 2|g| = 2)$ is always negative.

As a third example, we consider the case of $2|g| = 5$ [Fig. 4(c)]. In Table 3, we show all the combinations of $(N, N')$ but now in the range of $-6 \leq N, N' \leq 6$ and $|N - N'| \leq 4$. Using the symmetry of the Bessel function $J_{-n}(x) = (-1)^n J_n(x)$, we find in total 9 combinations giving non-zero contributions. When we take the terms of $|N - N'| = 0$ and 2, and neglect the terms containing $J_2$ and $J_{-2}$, which are relatively smaller than the other terms ($|J_2(5)| = |J_{-2}(5)| = 0.047$), we obtain

$$|T_{fi}^{\text{model}}(\boldsymbol{b} = 0, 2|g| = 5)|^2 \approx \sum_{N=-6}^{+6} |J_N(5)|^2$$

$$+ 8\exp\left(-\frac{\delta k}{k_e \sigma_\theta^2}\right)\{J_1(5)J_3(5)\cos(8\omega_{\delta k} t_p) + J_3(5)J_5(5)\cos(16\omega_{\delta k} t_p) + J_4(5)J_6(5)\cos(20\omega_{\delta k} t_p)\}$$
(A63)

and

$$M_{mn}^{\text{model}}(\boldsymbol{b} = 0, 2|g| = 5) = C_1 - C_2 - C_3 - C_1\cos(8\omega_{\delta k} t_p) + C_2\cos(16\omega_{\delta k} t_p) + C_3\cos(20\omega_{\delta k} t_p)$$
(A64)

with $C_1 = -8 J_1(5)J_3(5)\exp\left(-\frac{\delta k}{k_e \sigma_\theta^2}\right) > 0$, $C_2 = 8 J_3(5)J_5(5)\exp\left(-\frac{\delta k}{k_e \sigma_\theta^2}\right) > 0$, and $C_3 = 8 J_4(2)J_5(2)\exp\left(-\frac{\delta k}{k_e \sigma_\theta^2}\right) > 0$. The fastest oscillation period is $2\pi/(20\omega_{\delta k}) = \pi\, t_{\text{bunch}}$. Because the coefficient of $\cos(8\omega_{\delta k} t_p)$, $-C_1$, is negative while the other two coefficients $+C_2$ and $+C_3$ are positive, the modulation $M_{mn}(\boldsymbol{b} = 0, 2|g| = 5)$ can be positive. A positive peak is expected to appear at $t_p = 1/2 \times 2\pi/(8\omega_{\delta k}) = 3.9 t_{\text{bunch}}$, which is consistent with the result in Fig. 4(c).

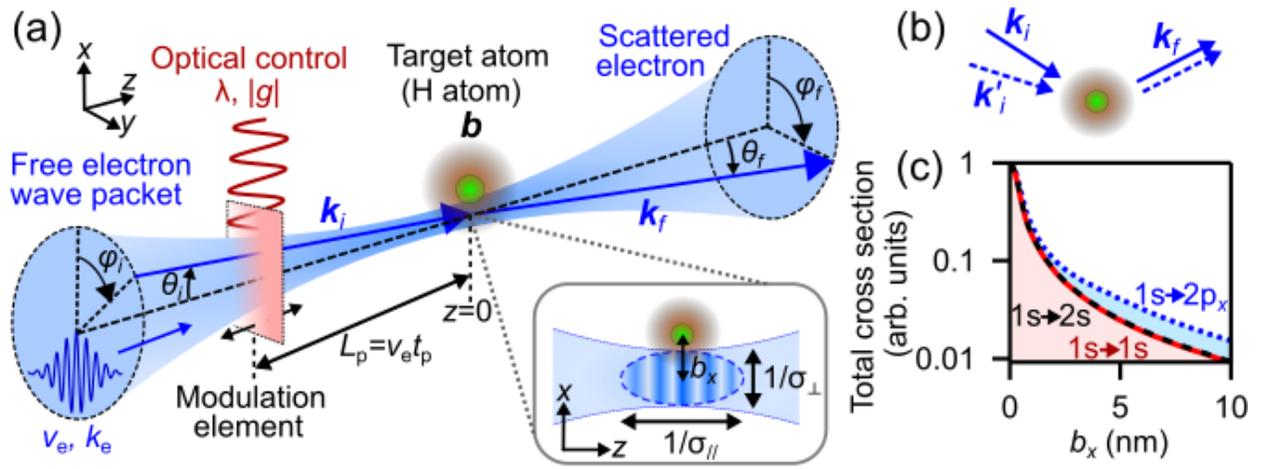

Fig.1 [single column figure]. Concept and scattering geometry. (a) A free electron wave packet (blue) is modulated longitudinally by an optical field (red) with a modulation element such as a membrane or a nanofabricated dielectric, depicted as a square. After the free-space propagation of distance $L_p$ and duration $t_p$, the electron wave packet is scattered by a field-free target (a hydrogen atom in the present study). The target is located at the spatial focal plane of the pulsed electron beam ($z = 0$). (b) Mechanism of quantum interference. Different momentum components of the incident electron wave packet ($\boldsymbol{k}_i$ and $\boldsymbol{k}_i'$) contribute to the scattering probability to the same final momentum ($\boldsymbol{k}_f$), which induces the interference. (c) The total scattering cross sections as a function of the impact parameter $b_x$ at $b_y = 0$. The cross sections decrease rapidly with $b_x$.

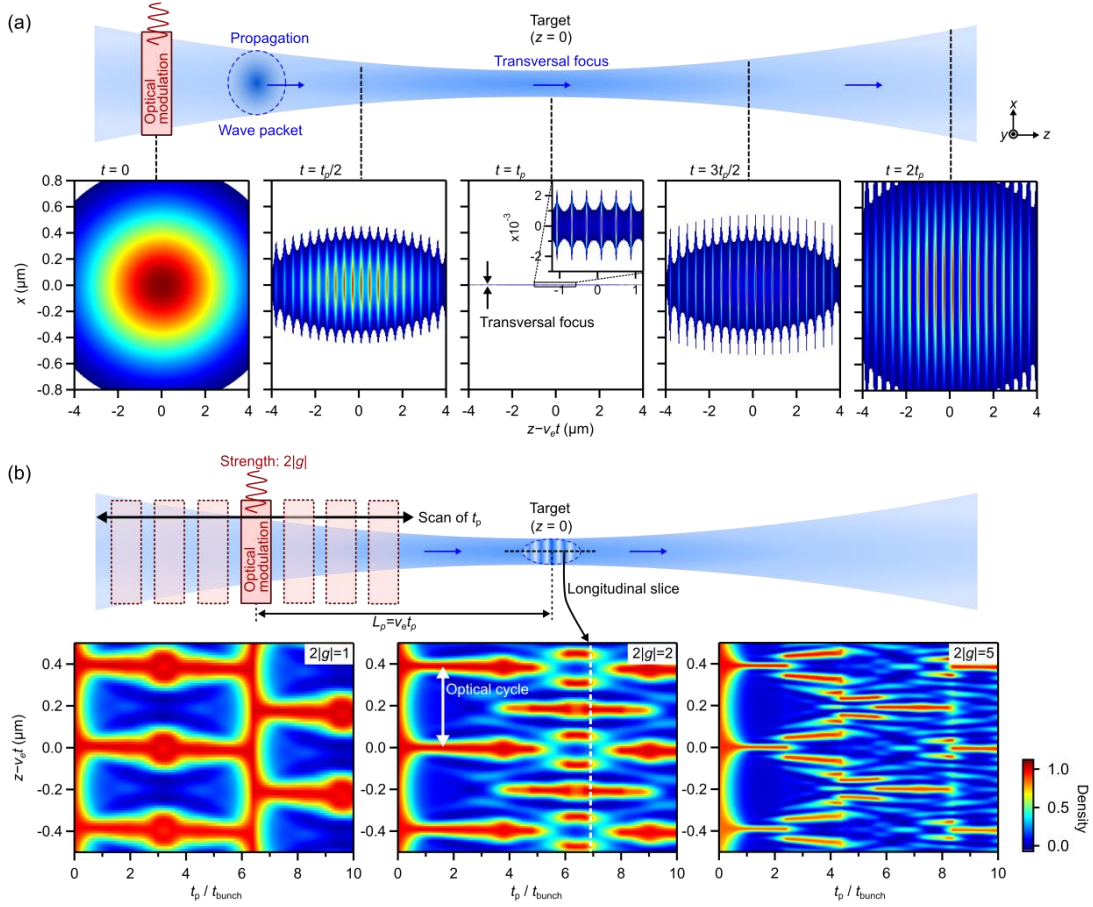

Fig. 2 [double column figure]. Free-space propagation dynamics of the electron wave packet $|\psi_e(x_e,t)|^2$ after the optical modulation. (a) Evolution of the optically-modulated 10-keV electron beam, calculated using Eqs. (4) and (21). At $t=0$ (left panel), the energy and longitudinal momentum distributions are modulated by a laser field of 2-µm wavelength but the real-space density has a Gaussian profile. After some free-space propagation ($t>0$), the longitudinal (temporal) density modulation can be seen. Simultaneously with the longitudinal density modulation, the transversal focusing/divergence occurs with the propagation. (b) Temporal density at the target position ($z=0$) at the modulation strengths of $2|g|=1$ (left panel), 2 (middle panel) and 5 (right panel), the three cases for the results in Fig. 4. Vertical profile at each $t_p$ corresponds to the longitudinal density profile. The propagation time $t_p$ from the optical modulation to the target, or equivalently, the location of the optical modulation, controls the temporal density and the scattering probabilities.

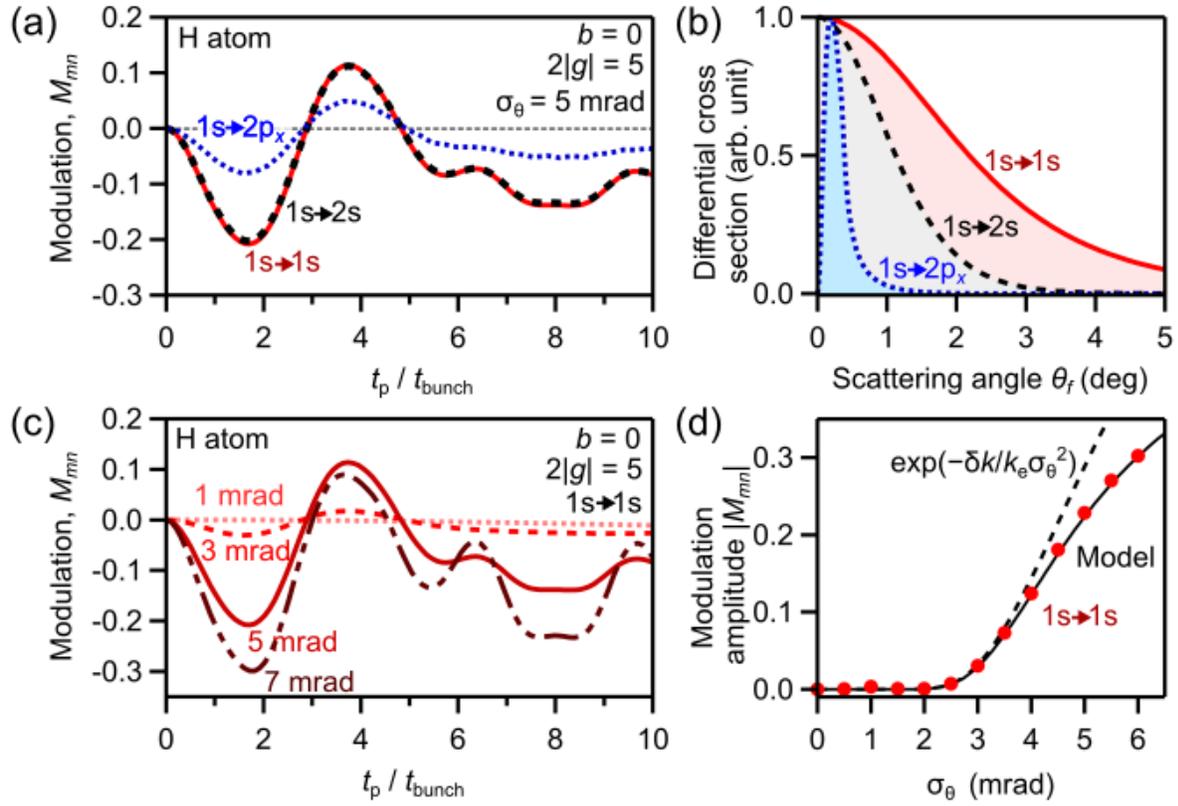

Fig. 3 [single column figure]. Modulation of total scattering probability at zero impact parameter, $M_{mn}(\boldsymbol{b}=0, g)$. (a) Comparison of three scattering processes. Qualitatively identical modulations are observed for these processes. (b) Differential cross sections for the three scattering processes. (c),(d) Beam focusing angle dependence. (c) Modulations of the total elastic scattering cross section for 4 different focusing angles. (d) Comparison with an exponential function (black dotted curve) and the simple model prediction (black curve). See text for details.

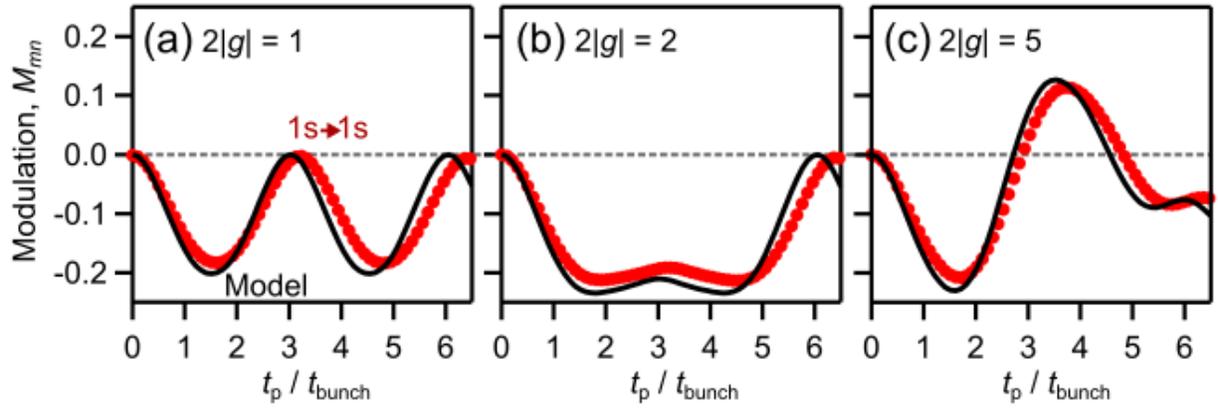

Fig. 4 [single column figure]. Modulation of the total elastic cross section (red circles) for $\boldsymbol{b} = 0$ at the optical coupling strength of $2|g| = 1$ (a), 2 (b), and 5 (c). The amplitudes and the shapes of the modulations are well reproduced by the simple model calculation (black curves). See text for details.

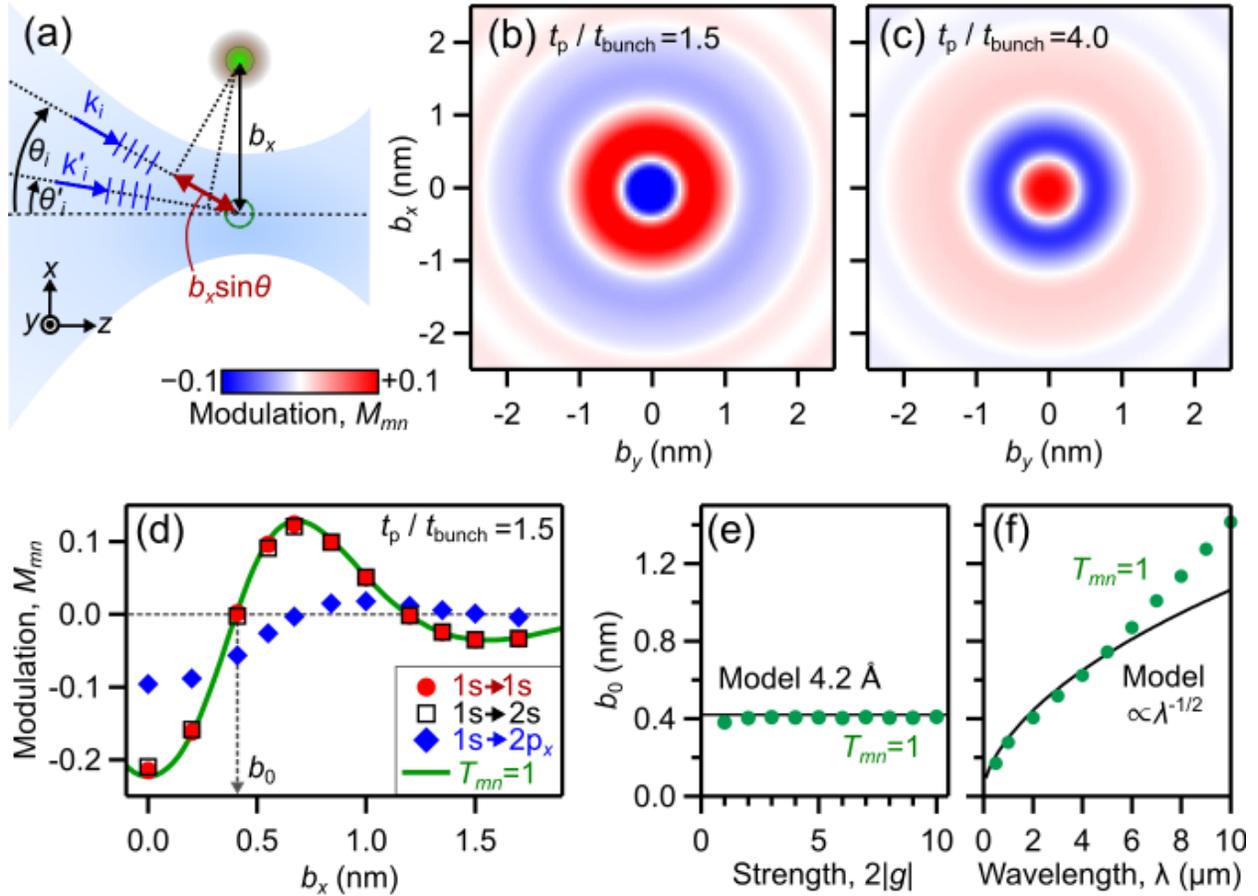

Fig. 5 [single column figure]. Non-zero impact parameter and the Angstrom-level spatial dependence. (a) Impact parameter and the associated phase. Depending on the angle $\theta_i$ ($\theta_i'$), a path length difference (red arrow) arises, which induces the relative phase of the two interfering components ($\mathbf{k}_i$ and $\mathbf{k}_i'$). The distance between the electron beam and the target is exaggerated for ease of illustration. (b),(c) Two-dimensional impact parameter dependence at $t_p = 1.5 t_{\text{bunch}}$ (b) and $t_p = 4 t_{\text{bunch}}$ (c) assuming $T_{mn} = 1$. (d) Slice at $b_y = 0$ and comparison with the three collisional processes of the hydrogen atom. (e),(f) Zero-modulation impact parameter ($b_0$) as a function of the optical modulation strength (e) and wavelength (f).

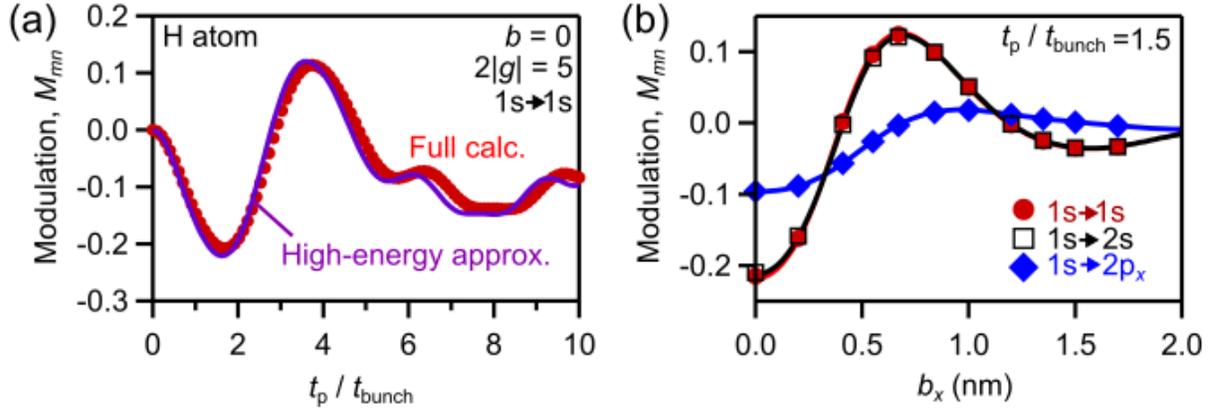

Fig. 6 [single column figure]. Validity of the high-energy approximation. (a) Simulation result for 10-keV electrons with the high-energy approximation with Eq. (34) (purple curve) compared to that of full simulation with Eq. (18) (red circles). The two results match well. (b) Impact parameter dependence at $t_p = 1.5 t_{\text{bunch}}$ for the three collisional processes of the hydrogen atom. Curves show the results of the high-energy approximation with Eq. (34). Circles, squares and diamonds show the simulation results with Eq. (18), the same as in Fig. 5(d). The results of the full simulations are well reproduced.

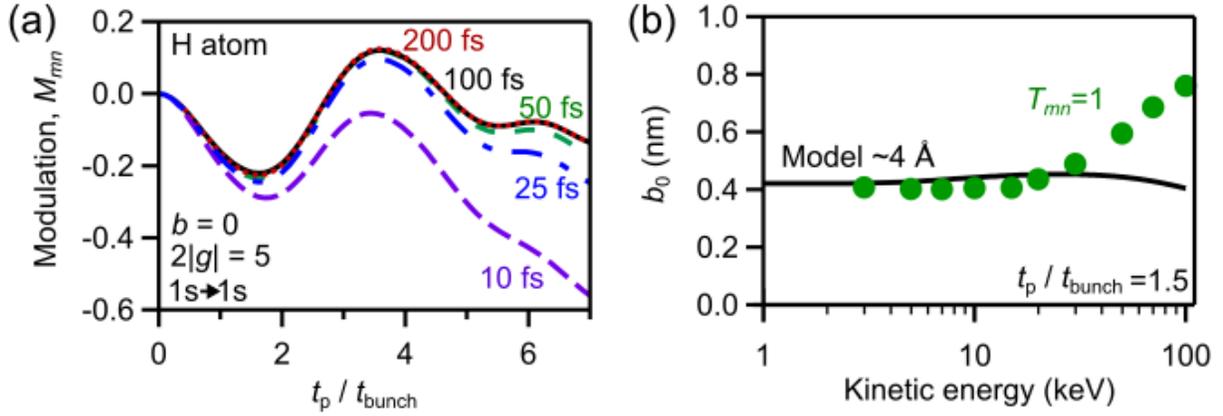

Fig. 7 [single column figure]. Beam parameter dependence. (a) Modulations $M_{mn}(\mathbf{b}=0, 2|g|=5)$ of total elastic scattering probability at 10 keV for five different longitudinal momentum widths $\hbar\sigma_\parallel$. The corresponding FWHM durations are shown. At durations longer than 50 fs, the modulation is nearly identical. At shorter durations, the dispersion of each $N$-photon component becomes significant and the deviations from the results with longer durations become noticeable at large $t_p$. (b) Zero-modulation impact parameter $b_0$ (see Fig. 5) as a function of the kinetic energy of the incident electron beam. Black curve shows the prediction of the simple model. Below 20 keV, $b_0$ is almost constant at $b_0 \sim 0.4$ nm, in agreement with the model prediction.

Table 1. Combinations of $(N, N')$ at $2|g| = 1$ with nonvanishing contribution to the scattering probability in the simple model.

| $|N^2 - N'^2|$ | $|N - N'|$ | $(N, N')$ | Net contribution |
|---|---|---|---|
| 1 | 1 | (1,0), (-1,0), (0,1), (0,-1) | 0 |
| 3 | 1 | (2,1), (-2,-1), (1,2), (-1,-2) | 0 |
| 3 | 3 | (2,-1), (-2,1), (-1,2), (1,-2) | 0 |
| 4 | 2 | (2,0), (-2,0), (0,2), (0,-2) | $4J_0(1)J_2(1) > 0$ |

Table 2. As caption of Table 1, but for $2|g| = 2$.

| $|N^2 - N'^2|$ | $|N - N'|$ | $(N, N')$ | Net contribution |
|---|---|---|---|
| 1 | 1 | (1,0), (-1,0), (0,1), (0,-1) | 0 |
| 3 | 1 | (2,1), (-2,-1), (1,2), (-1,-2) | 0 |
| 3 | 3 | (2,-1), (-2,1), (-1,2), (1,-2) | 0 |
| 4 | 2 | (2,0), (-2,0), (0,2), (0,-2) | $4J_0(2)J_2(2) > 0$ |
| 5 | 1 | (3,2), (-3,-2), (2,3), (-2,-3) | 0 |
| 5 | 5 | (3,-2), (-3,2), (-2,3), (2,-3) | 0 |
| 8 | 2 | (3,1), (-3,-1), (1,3), (-1,-3) | $4J_1(2)J_3(2) > 0$ |
| 8 | 4 | (3,-1), (-3,1), (-1,3), (1,-3) | $4J_{-1}(2)J_3(2) < 0$ |
| 9 | 3 | (3,0), (-3,0), (0,3), (0,-3) | 0 |

Table 3. As caption of Table 1, but for $2|g| = 5$ and $|N - N'| \leq 4$.

| $|N^2 - N'^2|$ | $|N - N'|$ | $(N, N')$ | Net contribution |
|---|---|---|---|
| 1 | 1 | (1,0), (-1,0), (0,1), (0,-1) | 0 |
| 3 | 1 | (2,1), (-2,-1),(1,2), (-1,-2) | 0 |
| 3 | 3 | (2,-1), (-2,1), (-1,2), (1,-2) | 0 |
| 4 | 2 | (2,0), (-2,0), (0,2), (0,-2) | $4J_0(5)J_2(5) < 0$ |
| 5 | 1 | (3,2), (-3,-2),(2,3), (-2,-3) | 0 |
| 7 | 1 | (4,3),(-4,-3),(3,4),(-3,-4) | 0 |
| 8 | 2 | (3,1), (-3,-1),(1,3), (-1,-3) | $4J_1(5)J_3(5) < 0$ |
| 8 | 4 | (3,-1), (-3,1),(1,-3), (-1,3) | $4J_{-1}(5)J_3(5) > 0$ |
| 9 | 3 | (3,0), (-3,0),(0,3), (0,-3) | 0 |
| 9 | 1 | (5,4), (-5,-4), (4,5), (-4,-5) | 0 |
| 11 | 1 | (6,5), (-6,-5), (5,6), (-5,-6) | 0 |
| 12 | 2 | (4,2), (-4,-2), (2,4), (-2,-4) | $4J_2(5)J_4(5) > 0$ |
| 15 | 3 | (4,1), (-4,-1), (1,4), (-1,-4) | 0 |
| 16 | 2 | (5,3), (-5,-3),(3,5), (-3,-5) | $4J_3(5)J_5(5) > 0$ |
| 16 | 4 | (4,0), (-4,0), (0,4), (0,-4), | $4J_0(5)J_4(5) < 0$ |
| 20 | 2 | (6,4), (-6,-4), (4,6), (-4,-6) | $4J_4(5)J_6(5) > 0$ |
| 21 | 3 | (5,2), (-5,-2),(2,5),(-2,-5) | 0 |
| 24 | 4 | (5,1), (-5,-1), (1,5),(-1,-5) | $4J_1(5)J_5(5) < 0$ |
| 27 | 3 | (6,3),(-6,-3),(3,6),(-3,-6) | 0 |
| 32 | 4 | (6,2),(-6,-2),(2,6),(-2,-6) | $4J_2(5)J_6(5) > 0$ |